\documentclass[amsfonts, amssymb, amsmath, reprint, showkeys, nofootinbib,superscriptaddress, aps, prl]{revtex4-2}

\usepackage{silence}
\usepackage{float}

\usepackage{graphicx}
\usepackage[left=15mm,right=15mm,top=25mm,columnsep=15pt]{geometry} 

\usepackage{adjustbox}
\usepackage{placeins}
\usepackage[T1]{fontenc}
\usepackage{csquotes}

\usepackage[nointegrals]{wasysym}
\usepackage{amsmath}
\usepackage{amssymb}
\usepackage{amsthm}
\usepackage{braket}
\usepackage{bm}
\usepackage{mathtools}
\usepackage{tensor}
\usepackage{mathrsfs}

\def\openone{\leavevmode\hbox{\small$1$\normalsize\kern-.33em$1$}}

\newcommand{\norm}[1]{\left\lVert#1\right\rVert}
 
\newcommand{\avg}[1]{\left< #1 \right>}

\let\baraccent=\= 
\renewcommand{\=}[1]{\stackrel{#1}{=}}

\newcommand{\be}{\begin{equation}}
\newcommand{\ee}{\end{equation}}

\newcommand{\ii}{\mathrm{i}}

\newcommand{\Tr}[1]{\textrm{tr}\left[#1\right]}

\usepackage[pdftex, pdftitle={Article}, pdfauthor={Author}]{hyperref}

\begin{document}
\title{Disorder-Induced Anomalous Diffusion in a 3D Spin Network}

  \author{Andrew Stasiuk}
    \email{astasiuk@mit.edu}
    \affiliation{Department of Nuclear Science and Engineering, Massachusetts Institute of Technology, Cambridge, Massachusetts 02139, USA}

  \author{Garrett Heller}
    \affiliation{Department of Physics, Massachusetts Institute of Technology, Cambridge, Massachusetts 02139, USA}
       
  \author{Lance Berkey}
    \affiliation{Department of Physics, Massachusetts Institute of Technology, Cambridge, Massachusetts 02139, USA}

  \author{Bo Xing}
    \affiliation{Research Laboratory of Electronics, Massachusetts Institute of Technology, Cambridge, Massachusetts 02139, USA}
    \affiliation{Quantum Innovation Centre, Agency for Science, Technology and Research, Singapore 138634, Singapore}
    
  \author{Paola Cappellaro}
    \email{pcappel@mit.edu}
    \affiliation{Department of Nuclear Science and Engineering, Massachusetts Institute of Technology, Cambridge, Massachusetts 02139, USA}
    \affiliation{Department of Physics, Massachusetts Institute of Technology, Cambridge, Massachusetts 02139, USA}

\date{\today} 

\begin{abstract}
Emergent hydrodynamics (EHD) bridges short-time unitarity with late-time thermodynamics, universal transport phenomena characterize the manner and speed of transport and thermalization. Typical non-integrable systems with few conserved local quantities are expected to be diffusive. In contrast, strongly disordered systems which admit phases such as many-body localization, are predicted to inhibit thermalization and thus slow dynamical transport. Disordered systems represent a uniquely poised platform to probe the quantum-to-classical transition and the emergence of irreversible thermodynamics from the underlying unitary structure. Here, we study a strongly disordered nuclear spin ensemble, using local measurements enabled by the disordered-state technique. We observe an apparent phase transition into a sub-diffusive regime, which we model as a random walk on the emergent fractal structure of a percolating network in the dipolar spin ensemble. Our novel theoretical model provides a framework for characterizing the emergence of thermalization in closed quantum systems, even in the presence of strong disorder.

\end{abstract}

\maketitle

\textit{Introduction -- } Emergent hydrodynamics (EHD), the study of how hydrodynamical equations and phenomena emerge from the microscopic details of interactions within many-body quantum systems, has garnered intense interest in recent years \cite{Castro2016, Gopalakrishnan2023, Bertini2021}. Hydrodynamical equations are able to describe the dynamics of complex systems with only a few degrees of freedom, serving as an intermediary description for an ergodic system before thermalization. Deriving macroscopic equations of motion from microscopic details plays a crucial role in understanding the emergence of the quantum-to-classical transition. The central quantities of interest are generally the form and speed of hydrodynamic transport. Namely, determining if motion is ballistic or diffusive, or neither, which is generically called ``anomalous diffusion''. Anomalous diffusion, as its name implies, is often \textit{a priori} unexpected, potentially emerging due to strong disorder or non-abelian Hamiltonian symmetries \cite{Kucsko2018, Ljubotina2019, Richter2021, Keenan2023}

Many of the theoretical studies of emergent hydrodynamics are confined to one-dimensional systems, where integrability yields analytic solutions that can be studied in the thermodynamic limit \cite{Castro2016}. Additionally, 1D systems can often be efficiently simulated numerically, with sufficiently large system sizes at sufficiently late times to reach the hydrodynamic regime. For example, matrix product state methods have been used to simulate large ($N \gtrsim 100$) strongly interacting quantum systems \cite{Ye2020}. Complete simulation of two-dimensional quantum systems at late times may be beyond reach \cite{Napp2022}, however approximate techniques, such as belief propagation, show promising progress towards efficient simulation of 2D systems \cite{Tagliacozzo2009,Guo2023}.

Experimentally, large-scale quantum simulators ($N \gtrsim 50$) with spatially resolved readouts have been very successful in probing emergent hydrodynamics, especially in 1D. Cold-atom and superconducting simulators have demonstrated an apparent phase transition to KPZ-like scaling at the Heisenberg point, where a non-abelian symmetry emerges \cite{Wei2022,Richter2021}. Recent advances in neutral-atom array and superconducting quantum devices have enabled the simulation of quantum matter in 2D \cite{Scholl2021, Ebadi2021, Su2023, Kim2023, Manovitz2024, Abanin2025}. While digital quantum simulators are capable of achieving high-dimensional connectivity in current devices \cite{Bluvstein2024}, the computational overhead limits the effective system size. High-dimensional digital quantum simulation will be of limited use until the advent of a scalable fault-tolerant device \cite{Preskill2018}. In the near term, there is a clear opportunity for native three-dimensional quantum systems to serve as programmable emulators of quantum matter. 

For this work, our programmable quantum emulator of choice is a solid-state NMR system. Understanding spin transport is crucial to predicting nuclear spin relaxation rates in solids, as well as facilitating dynamic nuclear polarization (DNP) \cite{bloembergen1949interaction, abragam1961principles, Abraham1959, Corzilius2020}. Direct measurement of spin-diffusion under the natural Hamiltonian has previously been enabled by high-fidelity decoupling sequences and strong magnetic field gradients \cite{zhang1998first}. More recently, it has been demonstrated that internally generated disordered fields can also provide access to transport properties via the local-autocorrelation even under engineered interactions \cite{Peng2023, martin2022localtherm}. The disordered field also provides access to interesting dynamical regimes, for example, sufficiently strong disorder can break the ergodicity of a nuclear spin ensemble, preventing efficient transfer of polarization during DNP\cite{DeLuca2015, DeLuca2016}.

In this Letter, we exploit the disordered-state technique, coupled with precise Hamiltonian engineering, to explore a novel regime of disordered matter in a 3D spin system with strong interactions. By varying the disorder strength and measuring the dynamical exponent, we find clear evidence of rapid slowing in the quantum transport process -- namely, anomalous sub-diffusion. To explain our results, we model the transport of magnetization as a random walk on a spin-network, where disordered fields can inhibit transport via edge cutting. This network undergoes a percolation phase transition at finite disorder which restricts the quantum walkers to a locally fractal geometry, inducing sub-diffusive behavior. Due to the simultaneous presence of quenched and annealed disorder, the static-lattice model fails to predict the precise dynamical exponent, showing a strong deviation from the Alexander-Orbach (AO) prediction. This model, and the absence of localization over the explored range of disorder strengths, may provide crucial insights into the fate of MBL and disorder-broken ergodicity in three-dimensional quantum systems \cite{Abanin2019}. These results have broad impacts on the potential efficacy of DNP-enhanced or MBL-protected quantum devices in the presence of strong disorder \cite{Huse2013, DeLuca2015}.

\textit{Methods -- } We work on an ensemble of strongly interacting fluorine nuclear spins, realized via NMR of a single-crystal sample of lithium fluoride placed in a large ($\sim$ 7.1 T) magnetic field aligned along the [111] axis. This orientation minimizes the nearest-neighbor coupling strength, which improves the Hamiltonian engineering fidelity at fixed Rabi power and pulse spacing. All twelve homonuclear nearest-neighbors of a fluorine spin have equal coupling strength in this orientation, $J_{nn}\approx 2\pi \times 4.6$ kHz.

Within the rotating frame generated by the magnetic field, the fluorine nuclear spin Hamiltonian has three main components: dipolar interactions, locally disordered $z$-field, and collective control.
\begin{equation}\label{eqn.hamiltonian}
\mathcal{H}(t) = \mathcal{H}_{D} + \mathcal{H}_{rZ} + \mathcal{H}_c(t),
\end{equation}
The collective resonant control allows for robust Floquet engineering of the homonuclear dipolar interaction and heteronuclear dipolar-generated disordered field:
\begin{align*}
\mathcal{H}_{D} &=\sum_{i<j} J_{ij}\left[\hat{S}_z^{(i)}\hat{S}_z^{(j)} - \frac{1}{2}(\hat{S}_x^{(i)}\hat{S}_x^{(j)}+\hat{S}_y^{(i)}\hat{S}_y^{(j)})\right]\\
\mathcal{H}_{rZ} &= \sum_{i} \omega_i \hat{S}_z^{(i)}.
\end{align*}
where $S_\alpha$ are spin 1/2 operators. Here, $J_{ij} = J_0 (1-3\cos^2\theta_{ij})/\norm{\bm{r}_i-\bm{r}_j}^3$, where $J_0 = (2\pi) \, 106.4$ kHz \r{A}$^3$ and $\theta_{ij}$ is the polar tilt of the displacement vector relative to the magnetic field. The disordered field is generated by the bath of thermal lithium spins, and is locally well described by a Gaussian distribution, $\omega_i \sim N(0,\sigma_0^2)$, with $\sigma_0 = (2\pi) \, 4.7$ kHz. The structure of the quenched lithium disorder is confirmed experimentally, with details given in the Supplemental Materials. Other sources of disorder in this system may stem from material defects, such as F-centers \cite{Klempt2003} and stray/inhomogeneous static fields. Further, as will be discussed later, the homonuclear Ising term in the interaction Hamiltonian also acts as a source of annealed disorder for the transport-mediating flip-flop interaction.

\begin{figure*}[t]
    \centering   
\includegraphics[width=1.00\textwidth]{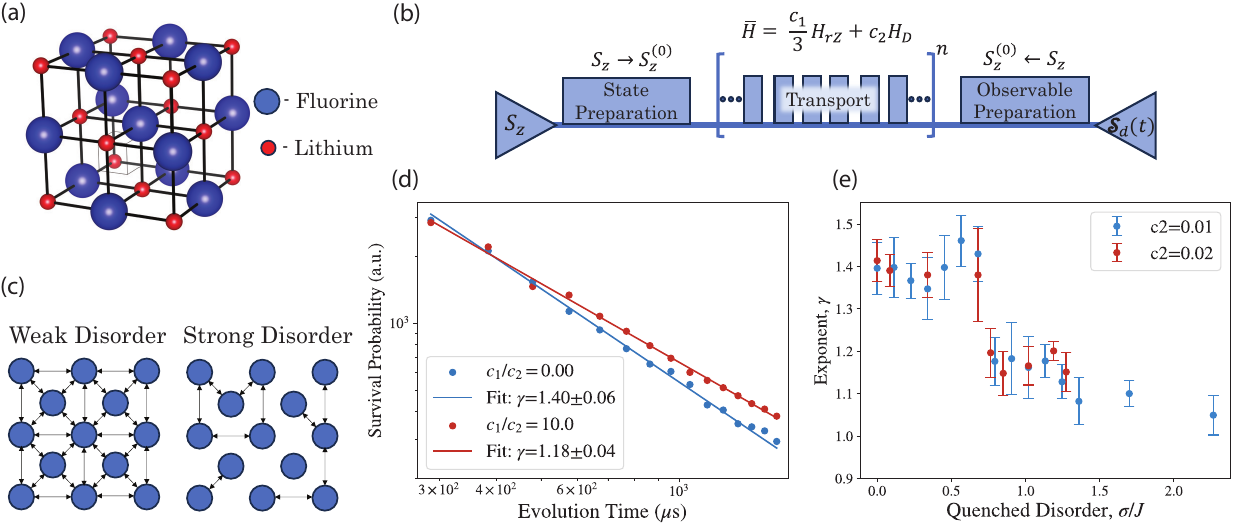}
    \caption{Here we show some of the details of the experimental system, and the collected hydrodynamic data. In panel (a), graphically depict the lithium (red) fluoride (blue) crystal structure. Each lithium spin is contributing to the disordered field of many fluorine spins, leading to spatial correlations with a $1/r^3$ profile. In panel (b), we provide a graphical depiction of our experimental sequencing -- a carefully engineered unitary transport block sandwiched by disordered state and observable blocks. Below the crystal structure, in panel (c), is an illustrative cartoon of the two conjectured hydrodynamic phases -- fully all-to-all connected in the weak disorder limit, and very sparsely connected in the strong disorder (fractal) limit. Due to the limited connectivity, large detours serve as deep traps and slow transport into a sub-diffusive phase. In panel (d) we plot two illustrative time-traces: one diffusive and one sub-diffusive. Finally, in panel (e), we plot the fitted survival probability exponent, $\gamma$, as a function of the quenched disorder strength generated by the Lithium spins, in units of $J_{nn}$.}
    \label{fig1}
\end{figure*}

By appropriate phase cycling while evolving under only the disordered field for sufficiently long times, we are able to generate states and observables with spatially uncorrelated phase-tagging, providing access to single-site readout. This technique has been previously demonstrated within fluorapatite \cite{Peng2023,Stasiuk2023}. Adapting the disordered-state technique to LiF required reducing the control error of the disordered field winding subroutine, which we achieved with the introduction of a new Hamiltonian engineering sequence. The details of this sequence, and its error scaling, can be found in the Supplemental Materials. Applying the control protocol depicted in Figure \ref{fig1} to an initially thermal state, we measure
\begin{equation}\label{eqn.local_auto}
    \mathcal{S}_d(t) = \sum_i \Tr{\hat{S}^{(i)}_z(t) \hat{S}_z^{(i)}} \propto \Tr{\hat{S}^{(0)}_z(t) \hat{S}_z^{(0)}}
\end{equation}
During the time $t$, the system evolves under an engineered target Hamiltonian consisting of dipolar interactions and a locally disordered field.  The relative magnitude of interactions to disorder is carefully controlled using the Wei16 Hamiltonian engineering sequence \cite{Wei2018}, yielding a two-parameter Hamiltonian:
\begin{equation}\label{eqn.transport_ham}
    \mathcal{H} = \frac{c_1}{3}\mathcal{H}_{rZ} + c_2\mathcal{H}_{D}.
\end{equation}
The two parameters, $c_1$ and $c_2$, rescale the strength of the quenched disordered field and dipolar interactions, respectively. 

Under the engineered interactions, the local autocorrelation signal, Eq.~\eqref{eqn.local_auto}, is measured and fit to an algebraic decay, $f(t) = \zeta t^{-\gamma}$. This fitting function captures the decay of the survival probability, and hence is (asymptotically) a witness to the speed and form of quantum transport. The value of $\zeta$ encodes the (generalized) diffusion constant. Unfortunately, it is not straightforward to normalize the experimental signal and thus extract the diffusion constant from this fitted value, as has been done in previous NV experiments \cite{Zu2021}. On the other hand, $\gamma$ encodes the universality class of the emergent hydrodynamic motion, with $\gamma=3/2$ corresponding to diffusion in three spatial dimensions.

We believe decoherence is negligible (or refocused) during the experiment. Reorientation of the lithium spins is expected to induce an exponential decay of the otherwise algebraic signal, but this should occur on a timescale of tens of seconds \cite{Kucsko2018}. Indeed, there is little evidence in the data to support the addition of an exponential envelope into the fitting function, as demonstrated in Figure \ref{fig1} (d). Experimental errors are more evident when we try to turn off all terms in the Hamiltonian, as also observed elsewhere~\cite{Sanchez20}.  For small values of disorder ($c_1\approx0$), we expect to measure $\gamma=3/2$. At various interaction strengths ($c_2=.01,\,.02$), the survival probability decays slightly slower than expected, $\gamma \approx 1.4 \pm .05$, but largely is consistent with the expectation of diffusive motion. Longer evolution times should improve the estimate of $\gamma$, though this directly competes with increasing control error accumulation and a decaying signal-to-noise ratio.

\textit{Results -- } 
Our experimental data, summarized in Figure~\ref{fig1}, indicate a clear emergence of two dynamical regimes -- diffusive at low disorder, rapidly yielding to sub-diffusive behavior above a critical disorder strength. The power-law fits which indicate sub-diffusive transport do not appear to vary strongly with the applied quenched disorder strength, indicating a robustness in the sub-diffusive regime which has not been measured elsewhere. These two distinct regimes, separated by a sharp transition, hint at a phase transition in the hydrodynamic universality class. To understand the observed dynamics, we develop a semi-classical random walk model on a disordered lattice network. This model naturally predicts a percolation phase transition in the dipolar connectivity graph. Hence, we attribute the observed sub-diffusive transport to the sensitivity of spin excitations to the percolating fractal network \cite{Gefen1982, Alexander1982, Bouchad1989}.

Previous investigations have demonstrated that a sufficiently strong disordered field can lead to sub-diffusive spin transport~\cite{Kucsko2018,Shi2024}. In sparse 3D long-range systems, like NV ensembles, the apparent sub-diffusion is explained with ``rare-resonance counting'', predicting $\gamma \sim 1/W$, for disorder strength $W$ \cite{Kucsko2018}.  Notably, rare-resonance counting is restricted to the regime where disorder is stronger than the interactions \cite{Kucsko2018}. Hence, this model does not correctly predict $\gamma=3/2$ in the low-disorder regime. A similar sub-diffusive trend has also been demonstrated in a 1D superconducting qubit ladder, wherein the sub-diffusion is attributed to the onset of Stark MBL, which is expected at large disorder strengths \cite{Shi2024}.

We now introduce our random walk model which treats the disorder as a modifier of the spin-network structure, in order to explain the onset of subdiffusion for quantum walkers. We start with the dynamics of an arbitrary spin pair, such that the restricted two-site Hamiltonian is $\mathcal{H}_{h}=\frac{c_1}{3}(\omega_1S_z^1+\omega_2S_z^2)+c_2\mathcal{H}_d^{12}$. The probability of ``hopping'', e.g. a spin-exchange $\ket{\uparrow \downarrow} \leftrightarrow \ket{\downarrow \uparrow}$, is a time-dependent function parameterized by the ratio of the random detuning and the strength of the coupling, $J_{12} = J_0 (1-3\cos^2\theta_{12})/\norm{\bm{r}_1-\bm{r}_2}^3$. The hopping probability for this spin-pair is thus $p_h= \left(1+\frac{4}{9}\frac{c_1^2}{c_2^2}\frac{(\omega_1-\omega_2)^2}{J_{12}^2}\right)^{-1}$. Since this is a random quantity, we also average over the disorder field distribution, $\delta\omega = \omega_1-\omega_2 \sim N(0,2\sigma_0^2)$, to find 
\begin{equation}
    \langle p_h\rangle=\frac{\sqrt{\pi } e^{1/\alpha ^2} \text{Erfc}\left(\frac{1}{\alpha }\right)}{2 \alpha },\qquad \alpha=\frac{4 c_1 \sigma_0}{3 c_2J_{12}}.
\end{equation}

We take the average hopping probability, $\avg{p_h}$, to be the bond acceptance probability for the spin network. In this manner, the disordered field modifies the connectivity of the spin network - highly detuned spin-pairs have their bonds ``cut'', disallowing exchange between them. When the disorder is very large, most bonds in the network will be cut. Conversely, at low disorder strengths, the network is fully connected. There is a critical disorder strength, $\sigma_c$, at which point the network percolates. At late times, and for fully quenched disorder, when $\sigma>\sigma_c$ the system is expected to fully localize. Otherwise, when $\sigma<\sigma_c$, diffusive motion is expected. Our experimental system, however, only probes finite times and is subject to a disordered field consisting of both quenched and annealed components. We estimate the critical disorder strength for the percolation phase transition by numerically computing the percolation strength and average cluster size in the quenched-only picture, the details of which can be found in the Supplemental Materials.

In order to efficiently predict the behavior of an $r^{-3}$-connected spin network, we also vary the bond connectivity, from nearest-neighbor (degree 12) up to sixth nearest-neighbor connectivity (degree 56). Here, we define the nearest neighbors to be the spins with maximal coupling strength. At the sixth nearest-neighbor level, we include bonds which are only $6.415\%$ of the nearest-neighbor strength. The extracted critical disorder strengths increase as a function of graph degree, from $\sigma_c\sim 3.2 J_{nn}$ for nearest-neighbor connectivity, up to $\sigma_c \sim 5.5 J_{nn}$ for sixth nearest-neighbor connectivity. Linear interpolation of this trend as a function of inverse graph degree predicts an infinite range critical disorder strength of $\sigma_c \approx 5.97 J_{nn}$.

At the percolation phase transition, the infinite connected subgraph is no longer Euclidean, rather, it has an emergent fractal structure. Away from criticality, the system still looks fractal at least up to the correlation length of the phase transition \cite{Poole1996, Kimmich2001}. Only sensitivity to length scales much larger than the correlation length will be perceived as Euclidean. For a classical walker, at asymptotically infinite time, motion is either diffusive or localized, except precisely at the phase transition point. However for finite time, which may still be many decades of time-steps, random walkers will perceive the lattice as being a fractal and transport sub-diffusively over a (possibly) wide range of disorder strengths. The survival probability of this walk would then decay as $t^{-\bar{\bar{d}}/2}$, where $\bar{\bar{d}}$ is the fracton dimension, introduced by Alexander \& Orbach, who found that $\bar{\bar{d}} \approx 4/3$ \cite{Alexander1982,Aharony1984}. In summary, our random walk model with quenched disorder predicts three distinct hydrodynamic regimes at finite times - diffusive, fracton sub-diffusive, and localized.

\textit{Discussion --} The experimental data from the lithium fluoride system is a clear indicator of the presence of two distinct hydrodynamic regimes -- diffusive and sub-diffusive. We confirmed the presence of this transition by considering two interaction strengths. While we were unable to explore sufficiently high disorder strengths which lead to localization in this work, previous studies on disordered NV center networks confirm our model's expected behavior in the localized regime \cite{Kucsko2018}. The strength of the lithium disorder inducing the phase transition is smaller than expected. We attribute this faster-than-expected motion to the presence of the homonuclear Ising term in the dipolar Hamiltonian, which can be viewed as an additional source of annealed disorder.  During transport, the local sampling of the homonuclear disordered field generated by this term will fluctuate at a rate which depends on the speed of the transport. A rapidly varying field can provide sufficient energy to tunnel through large energy barriers, pushing a disordered system towards the diffusive regime. As the quenched disorder increases in magnitude, the diffusivity will decrease, leading to slower variations in the annealed disorder, providing less energy for tunneling. The efficacy of the quenched disorder in inhibiting transport is coupled to the nature of the annealed disorder, which to our knowledge, has not been deeply investigated in any previous works.

The homonuclear Ising field is known to induce an effective detuning between interacting fluorine spins which depends on the magnetization of the ensemble \cite{Lowe1957, Greenbaum2005}. In previous hydrodynamic studies of dipolar ensembles, the homonuclear Ising term has been neglected \cite{Kucsko2018}, though it contributes significantly to the decoherence of multiple quantum coherence intensities \cite{Fedorov2006} and enters into numerical calculations of the magnetization diffusion constant \cite{Greenbaum2005}. The effect of this operator is well demonstrated by analyzing the transport term for two fluorine spins at sites $i$ and $j$. In the interaction picture of the homonuclear and heteronuclear Ising Hamiltonian, the transport operator transforms as
\begin{equation}\label{eq.homonuc_dis}
    \hat{S}_+^{(i)} \hat{S}_-^{(j)} \longrightarrow e^{\ii(\omega_i-\omega_j)t} \hat{S}_+^{(i)} \hat{S}_-^{(j)} \prod_{l\neq i,j}e^{2 \ii t (J_{li} - J_{lj})\hat{S}_z^{(l)}}.
\end{equation}
Whereas the lithium-disordered detuning $\omega_i-\omega_j$ is treated as independent and identically distributed quenched random variables, the homonuclear detuning term is not quenched (i.e., it is annealed) and depends weakly on the exchanging indices $i$ and $j$ due to spatial correlations in the homonuclear mean field. The experimental data shown in Figure \ref{fig1}  is evidence that the interplay between quenched and annealed disordered fields is non-trivial. In the sub-diffusive regime we measure $\gamma \approx 1.15$, strongly deviating from the classical AO prediction of $2/3$.

Due to the presence of the annealed disorder, the geometry of the connected fractal should rearrange itself in time while maintaining a well-defined mesoscopic fractal dimension. Monte Carlo studies of the homonuclear disordered field operator, given in Eq.~\eqref{eq.homonuc_dis}, indicate that the additional disorder is approximately Gaussian with zero mean. Therefore in addition to the variable quenched disorder $\sigma_{Li} = \frac{c_1}{3}\sigma_0$, maximally correlated nearest-neighbor exchange pairs feel a disorder strength of $\sigma_{F1} \approx 4c_2 \sigma_0$, whereas more distant uncorrelated exchange pairs feel a further magnified disordered field of $\sigma_{F2} \approx 5.5 c_2 \sigma_0$. In a 3D ensemble, the uncorrelated bulk is expected to dominate over the local neighborhood structure, hence the spatially averaged disordered field should skew closer to the uncorrelated $\sigma_{F2}$ value.

In the hydrodynamic regime, the state of the system is nearly thermal, hence the statistics of the homonuclear disordered field do not vary during transport. At a fixed moment in time, we can treat the homonuclear and heteronuclear disorders as instantaneously quenched and determine the total strength of the disordered field. For a system with only quenched disorder, we numerically predict that a percolation phase transition occurs $\sigma_c \approx 5.97 J_{nn}$ for a $1/r^3$ dipolar network. Notice that this is remarkably close to the estimated instantaneous total disordered field strength at the phase transition point between the diffusive and sub-diffusive regimes, $\sqrt{\sigma_{F2}^2 + (0.75 \sigma_0)^2} = 5.66 J_{nn}$. Therefore, it is evident that we have probed a regime where the correlation length of the percolation phase transition is non-negligible and is strongly interacting with the annealed disorder to generate a novel phase of fractal matter.

\textit{Conclusion -- } In this work, we studied emergent hydrodynamics in a strongly disordered nuclear spin ensemble. By probing a wide range of disorder strengths, we discovered a novel feature in the structure of hydrodynamic transport -- from diffusive to sub-diffusive. These experiments were made possible thanks to the disordered state technique in a LiF sample. To model this phase transition, we cast our hydrodynamic system as a continuous-time random walk on a percolation network, attributing the sub-diffusive behavior to the emergent fractal structure of the phase transition. Future probes of information scrambling and thermalization in this sub-diffusive regime would be extremely enlightening. Namely, we are curious whether this intermediate disorder inhibits ergodicity, or simply slows its onset. Clean theoretical models for predicting violations of ergodicity are vital to classifying whether quantum systems can thermalize under their interactions, which has a profound impact for engineered quantum devices, especially those relying on DNP \cite{DeLuca2015,DeLuca2016}. In addition,  within the sub-diffusive regime, we found an algebraic decay exponent of $\gamma\approx 1.15$, corresponding to a fracton dimension of $\bar{\bar{d}} \approx 2.3$, significantly higher than the Alexander-Orbach value of $\bar{\bar{d}} \approx 4/3$.

While there are existing theoretical models connecting Hilbert space fragmentation, percolation, and many-body localization \cite{Yoshinaga2022, Klocke2024, Kwan2025}, the percolation model proposed here is more easily verified in experimental devices thanks to its direct connection to hydrodynamic observables. Multiple quantum coherence experiments to determine coherence length should provide additional evidence of localization via direct measurement of the average cluster size. The feasibility of this experiment depends on the ability to reverse time evolution at late times. Additional theoretical studies in better understanding the perturbative mapping of unitary dynamics to a semi-classical percolation model could be very interesting, especially when exploring where the model breaks down. Due to strong finite size effects, it is unclear if direct numerical simulation can reach sufficient scales to capture the physics of the percolation phase transition.

Finally, we note that the implications of our percolation model are akin to that of Anderson localization. However, our Hamiltonian is strongly interacting, hence we cannot rule out the slow growth of entanglement entropy, e.g. many-body localization. Future experimental progress to definitively characterize whether the phase transition described here is consistent with higher dimensional MBL. It has been argued that MBL is thermodynamically unstable in high dimensional systems \cite{DeRoeck2017}; our subdiffusive transport regime may be direct evidence of the effects of that instability.

\textit{Acknowledgments -- } We thank Alex Ungar and Chandrasekhar Ramanathan for helpful hardware insights, Bob Kirby and Mitch Galaneck for providing time on the Gammacell Irradiator to engineer our sample, and Soonwon Choi for helpful discussions on the theory of many-body quantum dynamics. This work was supported in part by the National Science Foundation under Grants No. PHY1915218, No. PHY1734011, and No. OSI2326787. B.X. was supported by the A*STAR International Fellowship.

\bibliography{SDPrefs}


\end{document}


\widetext
\begin{center}
\textbf{\large Supplemental Materials: Disorder-Induced Anomalous Diffusion in a 3D Spin Network}
\end{center}

\setcounter{section}{0}
\renewcommand{\thesection}{S\Roman{section}}
\renewcommand{\thesubsection}{S\Roman{section}.\Roman{subsection}}
\makeatletter

\def\@seccntformat#1{\csname the#1\endcsname.\quad}
\@addtoreset{equation}{section}
\@addtoreset{figure}{section}
\makeatother
\renewcommand{\theequation}{S\Roman{section}.\arabic{equation}}
\renewcommand{\thefigure}{S\Roman{section}.\arabic{figure}}

\section{Hamiltonian Control}\label{sec:ham_control}
In our experiments, we utilized two central Hamiltonian engineering sequences, Wei16 and StaBerYXX. Wei16 is a 16-pulse sequence, composed of four similarly structured 4-pulse blocks \cite{Wei2018}. Unless otherwise specified, a pulse corresponds to a $\pi/2$ rotation about a particular transverse axis of the Bloch sphere obtained by turning on the rf driving for a time $t_p=1.1 \mu$s in our setup. Each 4-pulse block corresponds to 6$\tau_0$ of time, where $\tau_0$ is the delay primitive for the entire sequence. It is prudent to choose $\tau_0$ to be as short as possible in order to minimize Trotter errors (e.g., $\leq 5 \mu$s) while still ensuring interpulse delays $\tau-t_p$ are not too short. For our spectrometer, $\tau = 3.6\mu$s is as short as possible, given that the electronics require a minimum pulse separation of $2.5\mu$s to change the phase for the next applied pulse. 

Under Wei16, the internal dipolar Hamiltonian with disorder, given explicitly in the main text, can be transformed to a model $\mathcal H_F$ with 6 free parameters, 3 controlling the interaction Hamiltonian and 3 controlling the orientation of the local field: 
\begin{equation}\label{eqn.wei16}
    \mathcal{H}_F = \sum_{i<j} J_{ij} \big( (u-w)\hat{S}_x^{(i)}\hat{S}_x^{(j)} + (v-u)\hat{S}_y^{(i)}\hat{S}_y^{(j)} + (w-v)\hat{S}_z^{(i)}\hat{S}_z^{(j)}\big) + \frac{1}{3}\sum_i \omega_i\big(a\hat{S}_x^{(i)}+b\hat{S}_y^{(i)}+c\hat{S}_z^{(i)}\big).
\end{equation}
Notably, with collective rotations we cannot change the sum of the coefficients of the interaction Hamiltonian, since the subspace generated by the dipolar interaction along $x$, $y$, and $z$, written $\text{span}\lbrace \mathcal{H}_{D_x}, \mathcal{H}_{D_y}, \mathcal{H}_{D_z}\rbrace$, is a two-dimensional vector space, since $\mathcal{H}_{D_x} + \mathcal{H}_{D_y} + \mathcal{H}_{D_z} = 0$. Each of the parameters in Eq.~\ref{eqn.wei16}  are set by varying the physical pulse delays for the sequence,  
\begin{align}\label{eqn.delays}
    \nonumber \tau_1 &= \tau_0(1+c-v+w), \hspace{1cm} \tau_2 = \tau_0(1+b-u+v),\hspace{1cm} \tau_3 = \tau_0(1-a+u-w)\\
    \tau_1' &= \tau_0(1-c-v+w),\hspace{1cm} \tau_2' = \tau_0(1-b-u+v),\hspace{1cm}\tau_3 = \tau_0(1+a+u-w).
\end{align}
If $\tau_0$ is too short, then some of $\tau_i$ will be too short. Conversely, if $\tau_0$ is too large, then some of $\tau_i$ may be very long, and lead to large Trotter errors. For this work, $\tau_0=4\mu$s provides a good trade-off between dynamic range and error scaling.

In general, a four-pulse sequence is defined by five delays and four orientations, which we collect using the following notation, $P(t_1, \bm{n}_1, t_2, \bm{n}_2, t_3, \bm{n}_3, t_4, \bm{n}_4, t_5)$ \cite{Wei2018,Peng2023}. Unlike strings of unitary operators, this notation is read left-to-right, in the order they are physically applied during an experiment. Namely, the generic pulse sequence corresponds to ``wait for $t_1$, then apply a pulse in the $\bm{n}_1$ direction, then wait for $t_2$, ...'' and so on. Concretely, the Wei16 sequence is given by the string of four pulse blocks shown in the equation below:
\begin{align*}
    P(\tau_1,\bm{x},\tau_2,\bm{y}, 2\tau_3,\bm{y}, \tau_2',\bm{x},\tau_1') P(\tau_1',\bm{x},\tau_2,\bm{y}, 2\tau_3',\bm{y}, \tau_2',\bm{x},\tau_1) \cdots\\
    P(\tau_1,\bar{\bm{x}},\tau_2',\bar{\bm{y}}, 2\tau_3',\bar{\bm{y}}, \tau_2,\bar{\bm{x}},\tau_1') P(\tau_1',\bar{\bm{x}},\tau_2',\bar{\bm{y}}, 2\tau_3,\bar{\bm{y}}, \tau_2,\bar{\bm{x}},\tau_1)
\end{align*}
We use the standard notation $\bar{\bm{x}} = -\bm{x}$. The sequence has a great deal of reflection symmetry. Generally, symmetrized pulse sequences perform better than their unsymmetrized counterparts, which can be formalized by computing the leading-order error term.

\subsection{Random State Generation}\label{sec:random_state}
The procedure for generating the random state, and thus local observables in NMR, was first introduced by Peng \textit{ et. al.} on this experimental setup for a different crystal sample \cite{Peng2023}. Extensive details on the generation of random Zeeman states and random double-quantum states, along with discussion and verification of their properties, can be found in the supplementary material of that work. Here, we will reproduce some of the details relevant to our experiments, particularly the generation of random Zeeman states.

The goal is a unitary operator which maps our initial state to the disordered state,
\begin{equation}
    \hat{U}_d: \hat{S}_z \longrightarrow \sum_i \xi_i \hat{S}_z^{(i)},
\end{equation}
generated only using collective rotations. This is possible due to the local field generated by the heteronuclear interaction of our fluorine ensemble with nearby lithium atoms,
\begin{equation}
     \mathcal{H}_{dis} = \sum_{i,j} J^{FLi}_{ij} \hat{S}_z^{(i)}\hat{I}_z^{(j)} \approx \sum_{i} \omega_i \hat{S}_z^{(i)}.
\end{equation}
To make use of the disordered field, we must turn off the heteronuclear dipolar interaction while preserving the homonuclear dipolar interaction. This has been previously accomplished by using a symmetrized version of the WaHuHa sequence~\cite{Waugh68}.
\begin{equation}
    \text{WHH-8} =  P\big(\tau_0,\bm{x},\tau_0,\bm{y},2\tau_0,\bar{\bm{y}},\tau_0,\bar{\bm{x}},\tau_0\big)P\big(\tau_0,\bar{\bm{x}},\tau_0,\bar{\bm{y}},2\tau_0,\bm{y},\tau_0,\bm{x},\tau_0\big).
\end{equation}
For lithium fluoride, the error scaling of WHH-8 is unfavorable, as interactions are quite strong. In the LiF system, we replace WHH-8 with the StaBerYXX sequence, the details of which are provided in the next section.

The final ingredient needed to construct $\hat{U}_d$ is phase cycling, an NMR technique in which multiple experimental signals are added (or subtracted) together in order to cancel unwanted terms generated during evolution. All data within the main text used phase cycling to reduce systemic sources of noise generated by experimental imperfections and electronic control error. With this in mind, we present the procedure for generating random Zeeman states below.

\begin{enumerate}
    \item[(1)] The state is initialized via thermalization by waiting $5 T_1$, so that $\delta\rho_1 = \hat{S}_z$.

    \item[(2)] The state is rotated by a pulse along the $\bm{y}$ axis, so that $\delta\rho_2 = \hat{S}_x$. The state no longer commutes with the disordered field and will begin to evolve.

    \item[(3)] Immediately following the previous step, the StaBerYXX sequence is performed 5-7 times in a row, corresponding to $\tau_d = 12\tau_0 \times 5 \approx 200\mu$s of total evolution under the disordered Hamiltonian. The resulting state is then $\delta\rho_3 = \sum_i \cos(\omega_i\tau_d)\hat{S}_x^{(i)} + \sin(\omega_i\tau_d)\hat{S}_y^{(i)}$.

    \item[(4)] To get the disordered phase tagging along $\hat{S}_z^{(i)}$, a $\pi/2$ rotation is performed about the $\bm{x}$ or $\bm{y}$ axis. The choice is arbitrary as long as it is consistent; it will be clear that this choice determines whether the phase tagging is the sine or cosine of the disordered field. As an odd function, the choice of sine over cosine ensures the random phase tagging is zero mean even in the presence of finite temperatures. In this and previous works, $\bm{x}$ is chosen, leading to the state $\delta\rho_4^{(1)} = \sum_i \cos(\omega_i\tau_d)\hat{S}_x^{(i)} - \sin(\omega_i\tau_d)\hat{S}_z^{(i)}$. Two-fold phase cycling on the sign of the $\bm{x}$ pulse is performed. The first state is subtracted from another state generated by following steps (1)-(3), finally performing a $\bar{\bm{x}}$ pulse resulting in $\delta\rho_4^{(2)} = \sum_i \cos(\omega_i\tau_d)\hat{S}_x^{(i)} + \sin(\omega_i\tau_d)\hat{S}_z^{(i)}$. The subtraction of these states yields the desired result: $\delta\rho_4 = \delta\rho_4^{(2)}-\delta\rho_4^{(1)} \propto \sum_i\sin(\omega_i\tau_d)\hat{S}_z^{(i)}$.

    \item[(5)] Finally, a four-fold phase cycling is performed to reduce systematic error. The initial and final states resulting from steps (1)-(4) are entirely unchanged if \textit{all} pulses are shifted in phase by 90 degrees. Including this phase cycling step reduces error and leads to a larger signal, resulting in a procedure that requires an eight-fold phase cycling to generate $\hat{U}_d$.
\end{enumerate}

We have given a procedure for generating the random Zeeman state from the initial thermal state. The procedure to generate the random Zeeman observable is precisely the procedure used to generate the random Zeeman state, albeit in reverse. Following steps (1)-(5) in reverse will engineer $\hat{U}_d^\dagger$, leading to the desired observable. However, this procedure would require $8 \times 8 = 64$-fold phase cycling, so that each data point would take 1.5 hours to acquire. In engineering $\hat{U}_d^\dagger$, we omit the four-fold phase cycling detailed in step (5), so that we only need to perform a 16-fold phase cycle. This produces the same random-state observable albeit with less electronic error cancellation, which does not seem to significantly impact the measured signal. Shorter scans ensure that the spectrometer can be retuned often, ensuring maximal fidelity state preparation. For all experimental data shown in the text involving local observables, 16-fold phase cycling is used.

\subsection{Generating the StaBerYXX Pulse Sequence}\label{sec.staber_gen}

Many challenges in the experimental procedure needed to be overcome in order to successfully prepare disordered states in lithium fluoride. Firstly, it was quickly identified that WHH-8 had poor error scaling -- lithium fluoride is more strongly coupled and more highly connected than fluorapatite, leading to larger effective Trotter errors. Compensating for these errors with shorter Trotter steps was not possible on the existing hardware. Hence, we began exploring replacements for WHH-8 with a new spectral sequence in the disorder state generation subroutine.

In addition to investigating the use of Wei16 as a disorder generating sequence, we generated a family of new Hamiltonian engineering sequences for this purpose. Using the F-matrix formalism, new sequences can be rapidly and accurately prototyped. The F-matrix formalism, introduced by \cite{Choi2020}, collects the toggling frame operators of $S_z$, denoted $\lbrace \tilde{S}^z_i(t) \rbrace$, with respect to some Hamiltonian engineering sequence in a compact matrix form. In this way, straightforward algebraic relations are used to determine the sequence's error scaling. 

A Hamiltonian engineering sequence of length $k$ is thus defined by a $3 \times k$ shaped matrix, $F_{\mu i}$, and a length-$k$ time-bin vector $\mathbf{\tau}$. The length of $i$-th time delay is given by $(\mathbf{\tau})_i$. The representation of $S_z$ in the $i$-th toggling frame is specified by $\tilde{S}^z_i = \sum_{\mu} F_{\mu i} S_{\mu}$. If a sequence contains only integer multiples of $\frac{\pi}{2}$-rotations, then the F-matrix defines a series of permutations, such that each toggling frame operator is one of $\lbrace \pm S_\mu \,|\, \mu\in\lbrace x, y, z\rbrace\rbrace$. That is, $F_{\mu i} \in \lbrace0, \pm 1\rbrace$.

Evolution under the disordered field while decoupling interactions is required for generating the disordered state. There are many well-known pulse sequences for complete dynamical decoupling \cite{Cory1990}, as well as sequences which decouple homonuclear interactions while retaining heteronuclear interactions. Broadly speaking, these ``spectral sequences'' have been useful for molecular structure determination in solid state NMR. Because the target Hamiltonian is non-zero, it can be hard to adequately benchmark the sequence performance in a real sample \cite{Peng2022}.

Previous disorder state experiments on this hardware have utilized a symmetrized version of the WaHuHa sequence (WHH-8) to isolate the disordered field contribution \cite{Waugh68, Peng2023, Stasiuk2023}. Minimization of Trotter error and drift error requires minimizing the pulse-time $t_p$ and interpulse delay $\tau$. Pulse shape aberrations and electronic switching time requirements set hard minimums of $t_p=1.02\mu$s and $\tau=2.5\mu$s \cite{Stasiuk2024}. In practice, $\tau=3.6\mu$s is used. Decoupling of dipolar interactions can be done in $3\tau$ blocks known as solid echoes, though $6\tau$ sequences are required to ensure a return to the initial state. There are only 64 solid-echo-based $6\tau$ sequences (up to symmetries), none of which are directly suitable for enabling disorder-enhanced measurements as the disordered field component is always left along a titled axis. Ensuring the disordered field component lies along a single axis requires an 8-pulse $T=12\tau$ solid-echo based sequence.

\begin{figure}[h!]
    \centering
    \includegraphics[width=0.5\linewidth]{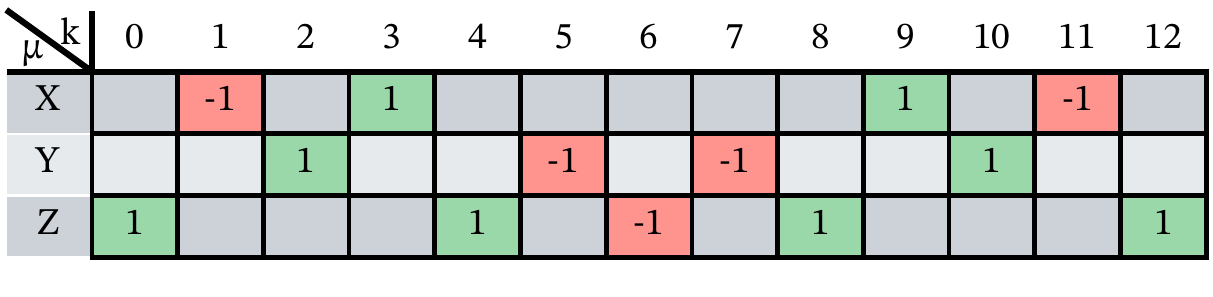}
    \caption{F-matrix of the StaBerYXX Pulse sequence. The delay vector length $k+1$ to include a half-delay at the begging and end of the pulse sequence, with timings $0.5\tau - \tau - \tau \dots -\tau - 0.5\tau$. Solid-echo based sequences have a stripped F-matrix structure, while YXX based sequences display chevron shaped structures, as seen here. The sequence is reflection symmetric over the central delay, which will innately cancel first order error terms.}
    \label{fig.s1}
\end{figure}

\begin{figure}[h!]
    \centering
    \includegraphics[width=0.5\linewidth]{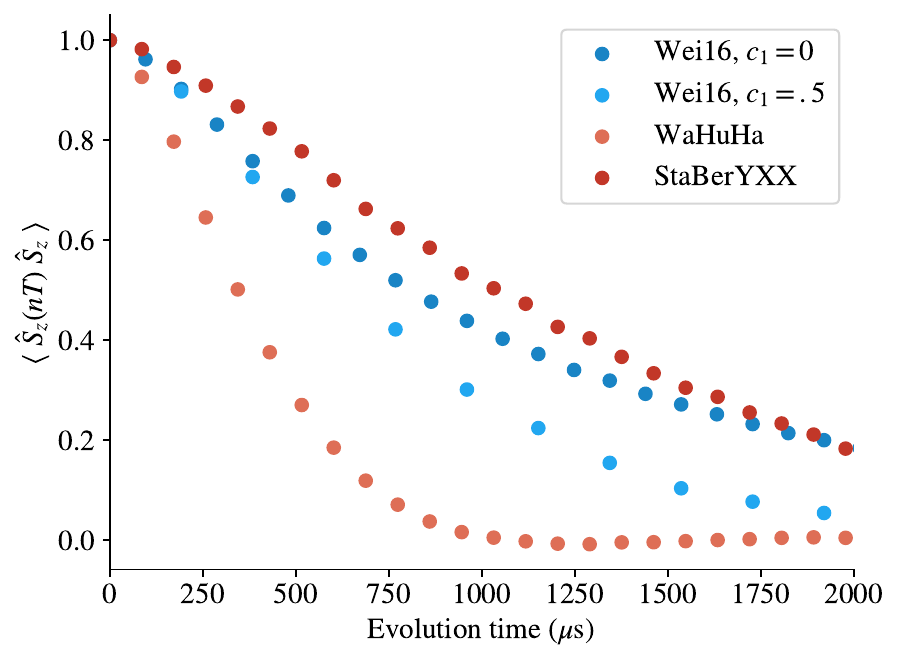}
    \caption{Decay of total magnetization under echo suspension experiment for each pulse sequence: StaberYXX, WHH-8, and Wei16. To first order, single-body terms will echo out, resulting in a 0 target Hamiltonian. Two-body error terms will not echo out and contribute to a non-zero effective Hamiltonian, resulting in decoherence in the global operator space. As expected from the theoretical analysis resulting in Table \ref{tab:first-order}, the StaBerYXX-based sequence boasts the slowest decay of fidelity during the suspension sequence.}
    \label{fig:suspension}
\end{figure}

To that end, we will restrict our search to $T=12\tau$ sequences. The WHH-8 sequence is a good example of an 8-pulse $12\tau$ sequence which leaves the disorder along $z$. However, the solid-echo structure results in a $\tau \,-\, 2\tau \,-\, \tau$ timing pattern, such that the $2\tau$ errors will limit the maximum achievable performance. In systems with fewer neighbors and weaker interactions, the $2\tau$ delay is not problematic and indeed it is often favorable to perform stroboscopic measurements in typical NMR experiments. In lithium fluoride, however, we experimentally verified (see Fig.~\ref{fig:suspension}) that WHH-8 is insufficient. Instead, we search over the space of $12\tau$ sequences generated from the YXX primitive, which has no $2\tau$ spacing requirement, consisting of 12 pulses \cite{Peng2022}. By restricting to a small search space, we are able to use F-matrix algebra to enable a brute-force search and find an optimal disorder generating pulse sequence. We refer to the resulting sequence as ``StaBerYXX'', and provide a graphical representation of its F-matrix in Figure \ref{fig.s1}.

We have three disorder generating sequences candidates:  WHH-8, Wei16, and our new sequence StaBerYXX. These sequences are all well-represented by the F-matrix, and thus can be compared against each other analytically. Denote $\overline{H}^{(0)}$ as the zeroth order term in the Magnus expansion given by average Hamiltonian theory. In terms of the toggling frame Hamiltonian between pulses, $\tilde{H}_k$,  
\begin{equation}
    \overline{H}^{(0)} =\frac{1}{T}\sum_{k} \tau_k \tilde{H}_k.
\end{equation}
Moreover let $\delta H_{avg}$ be the average drift error in zeroth order Hamiltonian due to finite pulses, up to leading order in $\frac{t_p}{T}$. The $k$-th control pulse is described by a unitary operator $U_k(t)$ generated by a time-dependent Hamiltonian, such $U_k(0)=\openone$ and $U_k(t_p)$ is a collective $\pi/2$-rotation. Then, $\delta H_{avg}$ can be written in terms of these ``partial pulse'' unitary operators,
\begin{equation}
    \delta H_{avg} = \frac{1}{T}\sum_{k} \int_{0}^{t_p} U^{\dagger}_{k}(t)\tilde{H}_k U_{k}(t) dt.
\end{equation}

 Both $\overline{H}^{(0)}$ and $\delta H_{avg}$ can then be computed with respect to the system Hamiltonian $\mathcal{H} = \mathcal{H}_{rZ} + \mathcal{H}_{Dz}$. Within the $i$-th toggling frame, $\tilde{H}_{rZ, k}$ and $\tilde{H}_{Dz,k}$ always mutually commute, but may have opposite signs. This can be seen since the dipolar interaction is invariant under collective $\pi$-rotations, whereas the disordered field may change sign. Then, the average Hamiltonian is simply the sum over toggling frame operators:
 \begin{align}\label{eqn:toggling_Z}
    \tilde{H}_{rZ, i} &= \sum_{\mu} F_{\mu i} \sum_{a} h_a S_\mu^{(a)}
\end{align}
\begin{align}\label{eqn:toggling_Dz}
    \tilde{H}_{Dz, i} &= \sum_{\mu} \left(\frac{3}{2}|F_{\mu i}| - \frac{1}{2}\right) \sum_{a,b} J_{ab}S_{\mu}^{(a)}S_{\mu}^{(b)}\\
    \nonumber \overline{H}^{(0)} &= \frac{1}{T}\sum_{i} \tau_i (\tilde{H}_{rZ, i} + \tilde{H}_{Dz, i}).
 \end{align}
 The results of these summations for all three sequences are reported in Table \ref{tab:hamiltonians}.
 
\begin{table}[h!]
    \centering
    \begin{tabular}{|c|c|}
        \hline
        Pulse sequence & $\overline{H}^{(0)}$ \\ \hline
        StaBerYXX & \( \frac{1}{6}\sum_i h_i S^z_i \) \\ \hline
        Wei16 disorder & \( \frac{1}{6}\sum_i h_i S^z_i \) \\ \hline
        WHH-8 & \( \frac{1}{3}\sum_i h_i S^z_i \) \\ \hline
    \end{tabular}
    \caption{Zeroth order ideal Hamiltonian for each pulse sequence. The zeroth order is computed by taking the time-weighted average of the Hamiltonian in each toggling frame, ignoring finite pulse effects. Each is proportional to $\mathcal{H}_{rZ} = \sum_{i} h_iS^z_i$, the disordered $z$ Hamiltonian. Indeed, all sequences are valid spectral sequences.}
    \label{tab:hamiltonians}
\end{table}

While all three sequences are mirror symmetric about their center, hence $\overline{H}^{(1)} = 0$, each of these sequences significantly differ in terms of their suppression of finite pulse errors. During a pulse, the Hamiltonian is in an intermediary position between $\tilde{H}_{k}$ and $\tilde{H}_{k+1}$. As such, there will be a cross term in the effective Hamiltonian between $\tilde{H}_k, \tilde{H}_{k+1}$ henceforth called $H^c_{k, k+1}$. The cross-term is computed in a straightforward manner using the parity-sum of the F-matrix, as previously derived in Choi \textit{et. al.} \cite{Choi2020}. During the brute-force search for StaBerYXX, we specifically only allowed sequences where the parity-sum operator evaluates to zero. In Wei16 the parity-sum operator also evaluates to zero.  However, the WHH-8 sequence has a non-zero parity sum operator, with cross-term
\begin{equation}
    \sum_{k}{H^{c}_{WHH, k, k+1}} =\frac{-6}{\pi}\sum_{ij}{J_{ij}}\left(S^x_{i}S^y_{j} +S^{y}_i S^{x}_j\right).
\end{equation}

The leading order drift error due to finite pulses $\delta H_{avg}$ is the sum over toggling frame operators, weighted by pulse duration, and the cross-term. Concretely,
\begin{equation}
    \delta H_{avg} = \frac{4}{\pi}\frac{t_p}{T}\sum_{k}{\tilde{H}_{rZ, k}} + \frac{t_p}{T}\sum_{k}{\tilde{H}_{DZ, k}} + \frac{t_p}{T} \sum_{k}{H^{c}_{k, k+1}}.
\end{equation}
The toggling frame operators have already been calculated in equations \eqref{eqn:toggling_Z} and \eqref{eqn:toggling_Dz}; it remains to re-sum them with the new weights. StaBerYXX is already equally spaced with no cross-term error, thus finite pulse error is parallel to the target Hamiltonian. On the other hand, WHH-8 and Wei16 have non-equally spaced pulses, so the pulse error may include an off-axis component in addition to the cross-term. Concretely, for Wei16 $\frac{t_p}{T}\sum_{ij}{J_{ij}\left(-2S_i^{x}S_j^x -2S_i^zS_j^z + 4 S_{i}^{y} S_{j}^{y} \right)}$ and $\frac{t_p}{T}\sum_{ij}{J_{ij}\left(-S_i^xS_j^x - S_i^zS_j^z + 2 S_{i}^{y} S_{j}^{y} \right)}$ for WHH-8, which are both simply a rescaling of the dipolar-y interaction Hamiltonian, $\mathcal{H}_{Dy}$. The disordered field modification due to finite pulse durations is $\frac{8t_p}{\pi T}\sum_{i}{h_i}{S_i^z}$ for both StaberYXX and WHH-8. The sum-totals with respect to our experimental parameters of $\tau=2.6$, $t_p=1.02$ are given in Table \ref{tab:first-order}. There are no chiral contributions to first order for any of the pulse sequences.

\begin{table}[h!]
    \centering
    \renewcommand{\arraystretch}{1.5} 
    \begin{tabular}{|p{3.5cm}|c|}
        \hline
        \textbf{Pulse sequence} & \textbf{Imperfect pulse Hamiltonian} \\ \hline
        StaBerYXX & \( H_{eff}^{(0)} = 0.18 \mathcal{H}_{rZ} \) \\ \hline
        Wei16 disorder & \( H_{eff}^{(0)} = 0.13 \mathcal{H}_{rZ} + 0.026 \mathcal{H}_{Dy} \) \\ \hline
        WHH-8 & \( H_{eff}^{(0)} = 0.33 \mathcal{H}_{rZ} + 0.013\mathcal{H}_{Dy} -0.049\sum\limits_{ij} J_{ij} ( S_i^y S_j^x +S_i^x S_j^y) \) \\ \hline
    \end{tabular}
    \caption{First order Hamiltonian for each pulse sequence, including leading order finite pulse errors. These are given by the average of the toggling frame Hamiltonian operators, along with the first-order cross-term errors. Of the three sequences, only StaBerYXX is proportional to $\mathcal{H}_{rZ} = \sum_{i} h_iS^z_i$, the disordered-$z$ Hamiltonian. Hence, StaBerYXX should be the most robust sequence for generating a disordered field along $z$ while decoupling all interactions.}
    \label{tab:first-order}
\end{table}

Based on the theoretical computations, we expect StaBerYXX to be the best spectral sequence, followed by Wei16 and WHH-8. WHH-8, though it has worse error scaling than Wei16, may still be useful as a spectral sequence due to the fact that WHH-8 is a $12\tau$ sequence, whereas Wei16 is $24\tau$. We confirm these error scaling predictions by experimental benchmarking the spectral sequences. 

It is much cleaner to characterize the performance of a sequence with target Hamiltonian of 0. Then, any decay of the signal must be due to errors. Notice, a static field can be echoed out with a $\pi$-pulse on a time scale of $2T$. Hence, we can test the performance of the ``echo suspension sequence'' generated by each of our spectral sequence candidates. A plot of magnetization as a function of evolution time for each sequence is shown in figure \ref{fig:suspension}. The StaBer-based sequence boasts a 2.67$\times$ and 1.43$\times$ increase in the $\frac{1}{e}$ survival time over the WHH-8 and Wei16 based sequences, respectively.

\begin{figure}
    \centering
    \includegraphics[width=0.5\linewidth]{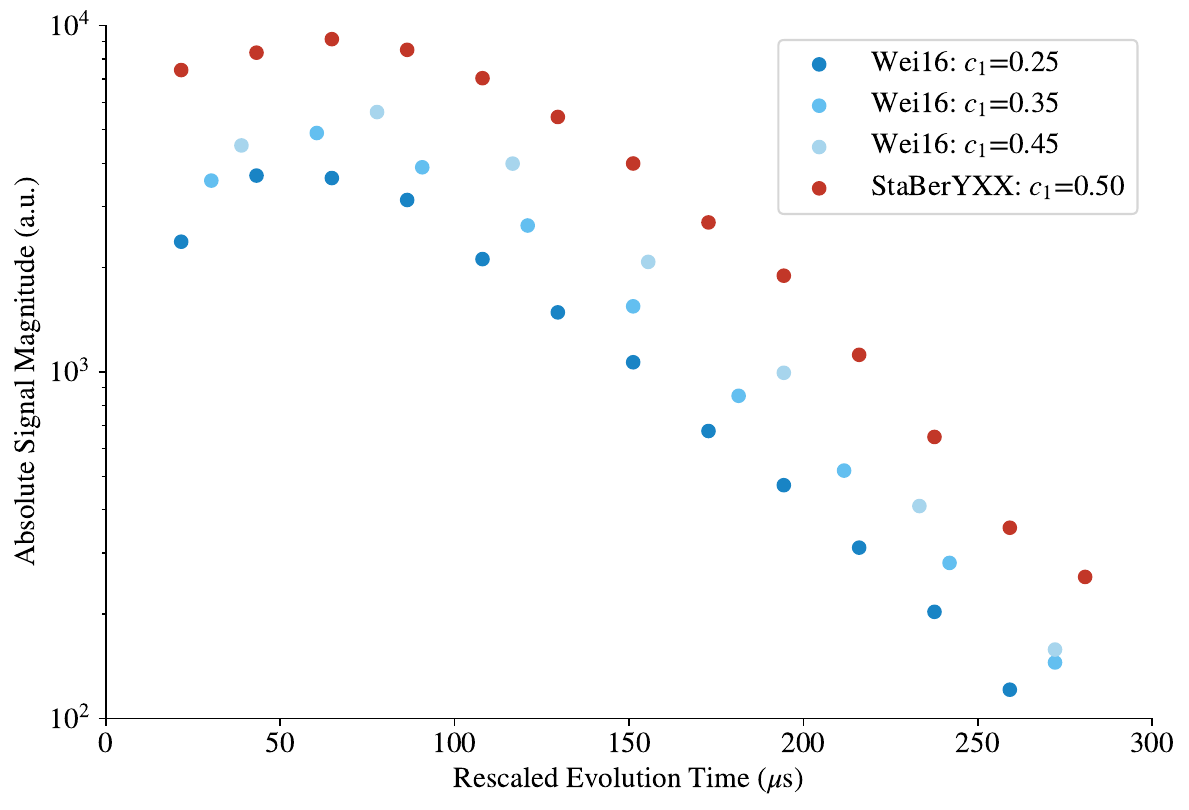}
    \caption{Disorder state contrast of each pulse sequence. Note the higher peak and tails of StaberYXX compared to Wei16 means we have a stronger signal to use in disordered state winding experiments when using StaberYXX.}
    \label{fig:disorder-fidelity}
\end{figure}

The other method of benchmarking performed here is based on the disordered-state generation protocol. Indeed, our motivation for studying high-fidelity spectral sequences is for use in the subroutines enabling disorder-assisted local measurements. After winding under the disordered field for $k$ Floquet cycles, with no Hamiltonian transport evolution, we would measure the following signal:
\begin{align}
    \nonumber \mathcal{S}(kT) &= \avg{ \sum_i \hat{S}_z^{(i)}(t) \sin \omega_i k T \sum_j \hat{S}_z^{(j)} \sin \omega_j k T}\\
    &= \sum_{i} \mathbb{E}\bigg[\sin^2 \omega_i k T\bigg] \Tr{\left(\hat{S}_z^{(i)}\right)^2}.
\end{align}
Here, $\omega_i \in \lbrace\frac{c_1}{3}h_i, \frac{1}{6}h_i\rbrace$ and $T\in\lbrace 12\tau, 24\tau\rbrace$, depending on the spectral sequence, as defined by the average Hamiltonian operators presented in Table \ref{tab:hamiltonians}. The absolute magnitude of this signal determines the base signal-to-noise ratio of the disordered-state protocol, and can be thought of as the ``disordered state contrast''. The amplitude of this signal should be maximal for optimal performance of the disordered state protocol. In Figure \ref{fig:disorder-fidelity}, this signal is plotted on a Log-Linear scale for various $c_1$ values of Wei16 and for StaBerYXX. After some initial rise time, all curves appear to decay exponentially with similar rates. Notably, the StaBerYXX sequence has the largest signal magnitude, and thus the best performance. In summation, the StaBerYXX sequence is a fast and highly performant spectral sequence, which enables the disordered state technique in a dense strongly-interacting dipolar spin ensemble. The measurements reported in the main text would not be possible using the WHH-8 subroutine, and severely limited by the Wei16 subroutine.

It is vital to maximize the signal-to-noise ratio for the disorder state preparation subroutine, since this procedure needs to last long enough to kill off long-range correlations. Concretely, after winding under the disordered field for $k$ Floquet cycles, and Hamiltonian evolution for some time $t$, we would measure the following signal:
\begin{align}
    \nonumber \mathcal{S}_d(t) = \avg{ \sum_i \hat{S}_z^{(i)}(t)\sin \frac{\omega_i k T_F}{6} \sum_j \hat{S}_z^{(j)} \sin \frac{\omega_i k T_F}{6}}\\
    = \sum_{i,j} \mathbb{E}\bigg[\sin \frac{\omega_i k T_F}{6}\sin \frac{\omega_j k T_F}{6}\bigg] \avg{\hat{S}_z^{(i)}(t) \hat{S}_z^{(j)}}.
\end{align}
At short times, this sinusoidal correlator inherits the $r^{-3}$ dependence of the bare disordered field. As will be discussed in Section \ref{sec:sample_char}, Monte Carlo simulations of 18,000 disorder-generating lithium spins show that this sinusoidal correlator collapses to a Kronecker-$\delta$ function after 75 $\mu$s of evolution under the bare disordered field. Namely, this collapse requires at least $k=5$ preparation cycles under StaBerYXX. The signal decays exponentially with increasing $k$ after this point, as seen in Figure \ref{fig:disorder-fidelity}, likely due to Trotter errors. For this reason, we fix to $k=5$ for most data collected for the main text. Experiments with up to $k=9$ cycles performed nearly identically in transport experiments, but with a reduced signal-to-noise ratio. Hence, $ \mathbb{E}\bigg[\sin \frac{5 \omega_i T_F}{6}\sin \frac{5 \omega_j T_F}{6}\bigg] \approx \delta_{ij}$, and so
\begin{equation}
    \mathcal{S}_d(t) = \sum_i \Tr{\hat{S}^{(i)}_z(t) \hat{S}_z^{(i)}} \propto \Tr{\hat{S}^{(0)}_z(t) \hat{S}_z^{(0)}}.
\end{equation}

If our preparation technique were to generate a short-ranged Gaussian profile instead of the assumed delta-function correlation profile, then this error would impact only the short-time power-law decay profile. The resulting signal would be a short-ranged integral of the diffusion kernel, such that $S_{err}(t) \sim \operatorname{erf}(1/\sqrt{t})/t$ rather than $S(t)\sim t^{-3/2}$. These curves are asymptotically equivalent, but differ strongly at short times. Short time errors are not an issue, since already the signal starts at finite thermal magnetization with a parabolic turnover before universal scaling dynamics take-over. There is no evidence that these early-time errors are strongly impacting the measured local autocorrelation signals.

\section{Sample Engineering and Characterization}\label{sec:sample_char}

All previous NMR studies utilizing the disordered state technique have focused on the fluorapatite sample, a quasi-1D nuclear-spin ensemble \cite{Peng2023, Stasiuk2023}. In the fluorapatite system, each computational fluorine spin is closely surrounded by three phosphorus spins. The phosphorus spins serve as an effective source of disorder, enabling the disordered state technique. In order to simulate 3D systems, we searched for a new fluorine-based crystal with a prevalent secondary nuclear spin-species for generating disorder. We settled on lithium fluoride, as it was both readily available and historically relevant in dual-species NMR literature \cite{abragam1961principles}.  

To homogenize nearest neighbor interaction strengths, and reduce the average coupling strength within the crystal, we oriented the crystal such that the magnetic field was aligned to the sample's [111] axis. Slowing the timescales of interactions improved our ability to analyze free induction decay (FID) measurements, thereby improving the quality of spectrometer tune-ups. The crystal orientation relative to the field is maintained by a 3D-printed a plastic cylinder with a depression of a cubic corner in its middle, so that a sample cut along the $x,y,z$ faces would stand on its point. We did not find that the plastic's hydrogen content significantly effected the bulk coherence properties of the lithium fluoride sample. However, it did meaningfully shift the effective dielectric constant of the induction coil for measurement and control, as compared to the paradigmatic glass sample-tube. This shift was compensated by modifying to the coil winding density of the probe inductor in order to more effectively tune the probe circuit to be resonant with the fluorine spins. The new sample holder might also impact the magnetic field homogeneity seen by the sample, due to the reduction in symmetry of the new sample holder. Line broadening due to field inhomogeneity was empirically a smaller effect than the line-narrowing due to the reduction in effective coupling strength in the [111] orientation compared to the [001] orientation.

To initialize to a thermal state, we at minimum must wait $5T_1$ between scans. Preliminary calibrations indicated that the sample $T_1$ was many tens of minutes long, causing even basic calibration experiments to be extremely onerous. Direct spin-lattice coupling of nuclear spins is quite small; solid-state NMR $T_1$ times are usually dominated by the sample's defect density \cite{abragam1961principles}. Deviations from magnetic equilibrium transport diffusively until colliding with electronic spin defects, which couple strongly to the lattice bath and serve as a local sink for magnetization \cite{abragam1961principles, Klempt2003}. Material defects can be introduced during crystal growth or created during an inelastic process leading to material damage. 

Under sufficiently high-energy electromagnetic irradiation, vacancy-interstitial pairs will form in the lithium fluoride lattice. The lattice vacancies can then form F-centers, a colored defect formed when an electron spin fills in the vacancy. The material can be healed with sufficiently high-temperature annealing, which allows for lithium fluoride to be used as a radiation dosimeter. More relevant to our goals, is that the F-center defects are paramagnetic, hence the system $T_1$ is a decreasing function of F-center density. We exposed the crystal to 5 MGy of $\sim$ 1 MeV $\gamma$-radiation from a Co-60 source to produce an abundance of paramagnetic F-centers. After the irradiation, our sample $T_1$ fell to $18.5$s (and the color changed from a clear yellowish to a deep red), consistent with expectations from previous NMR studies of irradiated lithium fluoride \cite{Klempt2003,Stork2008}. 

\section{Disorder field modeling and experimental calibration}
The controllable component of the disordered field is \textit{a priori} well described by a zero-mean normal distribution, similar to the disordered field seen in fluorapatite. Given substantial differences in the sample geometry, we reverify these claims for the lithium fluoride system. Recall that the lithium disorder is a consequence of the heteronuclear dipolar interaction,
\begin{equation}
     \mathcal{H}_{dis} = \sum_{i,j} J^{LiF}_{0} \frac{1-3\cos^2\theta_{ij}}{\norm{\bm{r}_i - \bm{r}_j}^3} \hat{S}_z^{(i)}\hat{I}_z^{(j)}. 
\end{equation}
We use $S$-labeled operators for fluorine nuclear spins, and $I$-labeled operators for lithium nuclear spins. When in thermal equilibrium at room temperature, the Zeeman polarization of the lithium spins, $\hat{I}_z^{(i)}$, is well approximated as a classical random variable which is uniform over all spin projections. The dominant isotope in natural abundance lithium fluoride is lithium-7, with $I=\frac{3}{2}$; at less than 5\% abundance, we will neglect the effects of lithium-6. Hence we replace the quantum operators $\hat{I}_z^{(i)}$ with a uniform random variable $I_z^{(i)} \sim \text{Uniform}(\pm\frac{1}{2},\pm\frac{3}{2})$. 

\begin{figure}[th]
    \centering
    \includegraphics[width=0.5\linewidth]{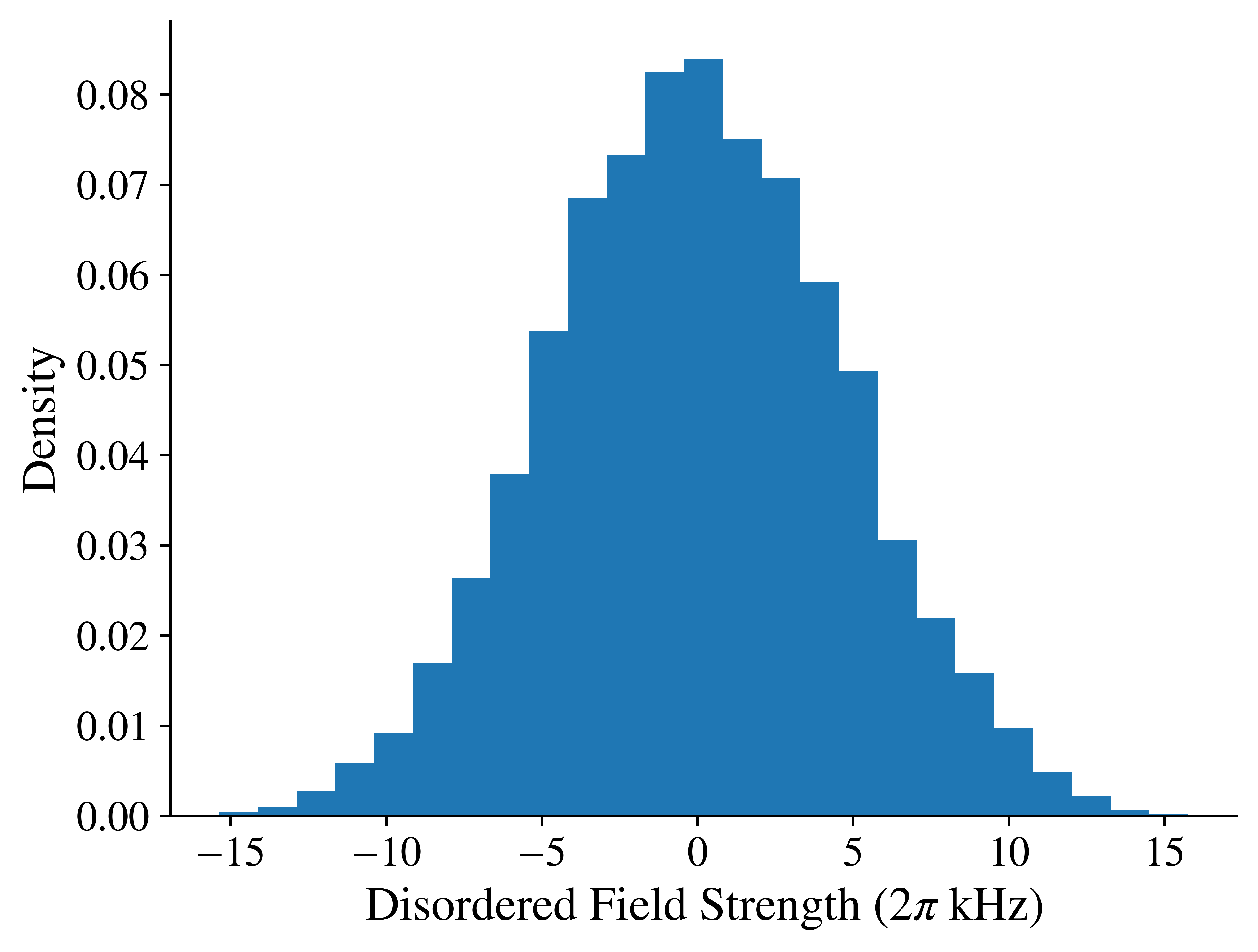}
    \caption{Disordered field distribution generated by 17,968 lithium spins in the LiF lattice, centered on the origin. The distribution is approximately normal with 0 mean and standard deviation of $\sigma_0 \approx (2\pi) \, 4.7$ kHz.}
    \label{fig.disorder_distro}
\end{figure}

In this way, we can write down a mean-field description of the lithium spins, as seen by the fluorine spins, such that
\begin{equation}
    \mathcal{H}_{dis} = \sum_i \gamma_F B_d(\bm{r}_i) \hat{S}_z^{(i)} = \sum_i \omega_i \hat{S}_z^{(i)}.
\end{equation}
By translational invariance of the crystal, it suffices to characterize $B_d(\bm{0})$ and its correlations. The statistics of the local disordered field can be numerically estimated either with direct summations or generating a histogram distribution by Monte-Carlo sampling of $I_z^{(i)}$, as previously demonstrated for fluorapatite \cite{Peng2023,Stasiuk2024}.

Figure \ref{fig.disorder_distro} shows 10,000 instances of the disordered field distribution generated by 17,968 lithium spins. A probability plot confirms that the generated distribution is approximately normal, with some deviations in the tails. Further, the distribution mean is approximately 0 as expected, with standard deviation $\sigma_0 = (2\pi)\, 4.702$ kHz.  We are also able to probe this value experimentally, to verify the estimated disordered field standard deviation. Via the Hamiltonian engineering sequences of MREV-8 and StaBerYXX, the system will evolve under only the disordered field, up to modifications of its strength and direction. MREV-8 tilts the disordered field and rescales its magnitude, such that $\bm{B} = B_d\bm{z}\longrightarrow \frac{B_d}{3}\big(\bm{x} + \bm{z}\big)$. The StaBerYXX sequence only rescales the field strength, $\bm{B} = B_d\bm{z} \longrightarrow \frac{1}{6}B_d\bm{z}$. 

\begin{figure*}
  \centering
  \includegraphics[width=0.48\linewidth]{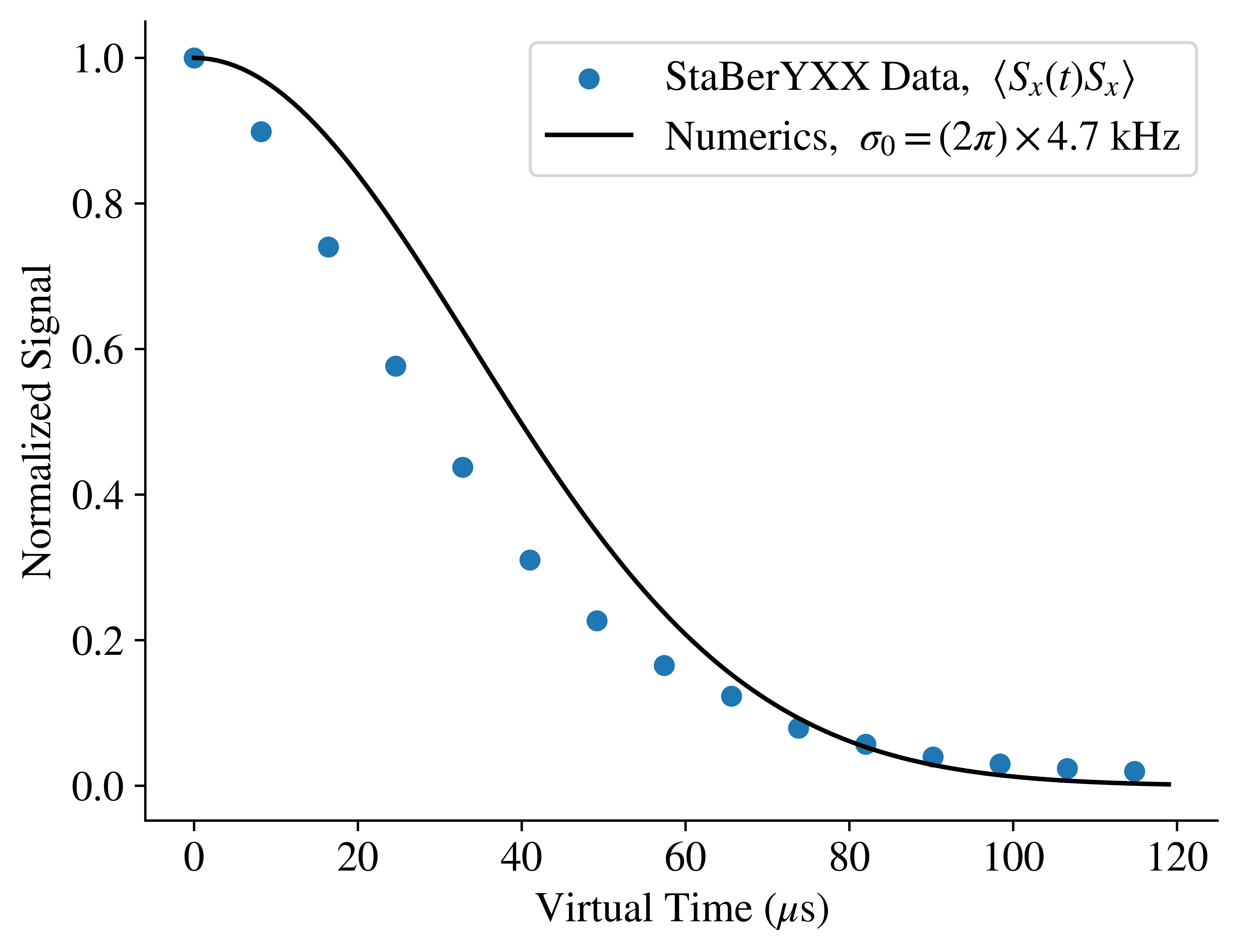}
  \hfill
  \includegraphics[width=0.48\linewidth]{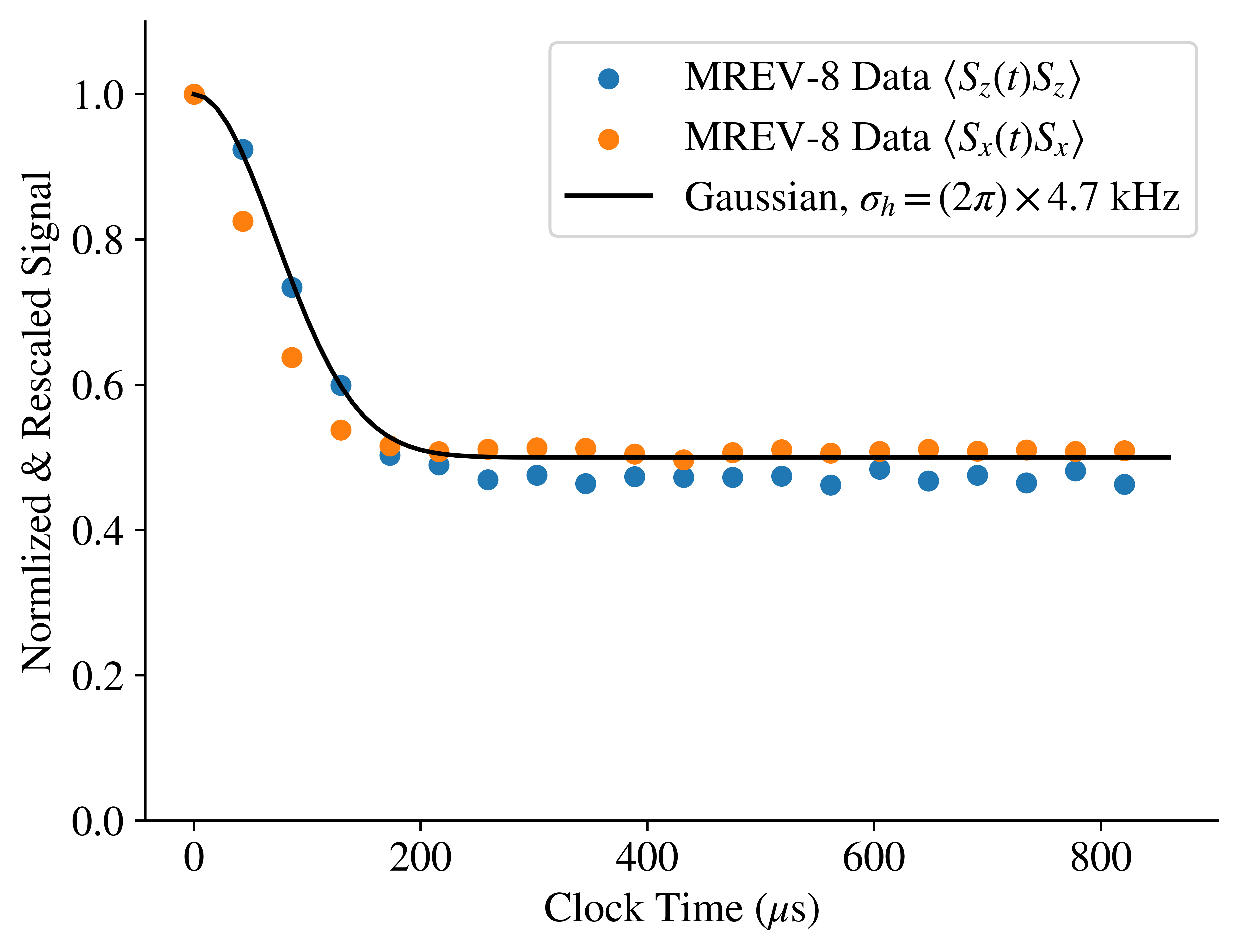}

  \caption{Experimental signals from evolution of global observables under the only the disordered field, using Hamiltonain engineering sequences StaBerYXX (left), and MREV-8 (right). The Staber signal is normalized to a maximum signal of 1, and the evolution time is rescaled by the disordered field prefactor of $1/6$. The MREV-8 signal is itself rescaled by an exponential decay, fitted from the decay of the last 15 data points, and is plotted relative to the clock-time of the experiments. Both sets of experimental data agree well with the numerical predictions for the disordered field distribution, presented in Figure \ref{fig.disorder_distro}. Both figures show that there may be slightly more disorder than expected (since the curves decay faster than the numerical prediction), though still largely consistent with the single-species lithium model. The StaBerYXX data is clearly inconsistent with a Gaussian line-shape, however we attribute this to the fact that we are unable to rescale by an exponential decay as was done with the MREV data.}
  \label{fig.disorder_cal}
\end{figure*}

Since homonuclear interactions are canceled under these spectral sequences, operator evolution can be analyzed analytically. The details of this analysis have been carried out previously \cite{Stasiuk2024}, such that for MREV,
\begin{equation}
    \overline{\avg{\hat{S}_z(t)\hat{S}_z}} = \overline{\avg{\hat{S}_x(t)\hat{S}_x}} = \frac{1}{2} + \frac{1}{2}e^{-\frac{1}{9}(\sigma_0 t)^2}.
\end{equation}
For the StaBer sequence, $\hat{S}_z$ commutes with the average Hamiltonian and should not evolve, while $\hat{S}_x$:
\begin{equation}
    \overline{\avg{\hat{S}_x(t)\hat{S}_x}} = e^{-\frac{1}{2}(\sigma_0 t/6)^2}.
\end{equation}
For the case of the tilted field, the late time behavior is expected to decay to a constant value of $1/2$, which is not seen experimentally due to pulse errors and other sources of decoherence. The deviation from this asymptote can be fitted to an exponential decay, and used to normalize the signal over the whole range of evolution. Such a rescaling is not possible for the StaBer sequence, even though the sequence itself is expected to be higher fidelity. The data for these experiments is shown in Figure \ref{fig.disorder_cal}, and has good agreement with the numerical predictions.

Disordered field correlations, if they are significantly long-ranged, can impact more than just the rate of emergent hydrodynamics, but also the universality class of motion \cite{Bouchaud1987,Bouchad1989}. The disordered field correlation can be expressed as a sum,
\begin{equation}
    \avg{B_d(\bm{r})B_d(\bm{0})} = \frac{5 (J_0^{LiF})^2}{4\gamma_F^2}\sum_j \frac{(1-3\cos^2\theta_{jr})(1-3\cos^2\theta_{j0})}{\norm{\bm{r}-\bm{r}_j}^3\norm{\bm{r}_j}^3} \sim r^{-3}.
\end{equation}
These correlations are critically long-ranged in 3D, but do not induce anomalous diffusion on their own \cite{Bouchaud1987}. Hence, while these correlations may impact the width of a percolation phase transition \cite{Argyrakis1985}, they are not significant enough to perturb the fundamental structure of our random walk model. For percolation numerics discussed in the next section, we treat the disordered field as effectively independent and identically distributed.

\begin{figure*}
  \centering
  \includegraphics[width=0.48\linewidth]{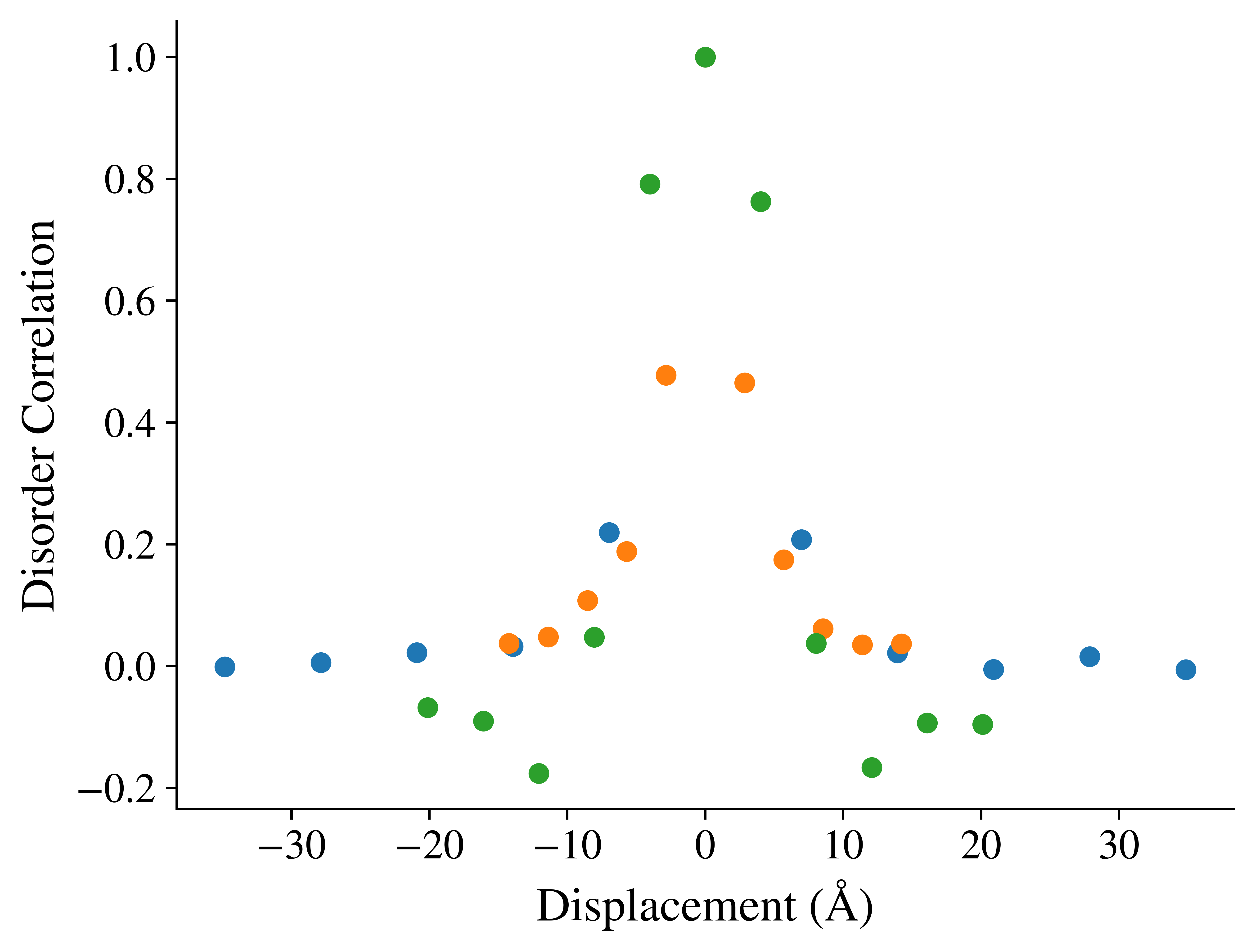}
  \hfill
  \includegraphics[width=0.48\linewidth]{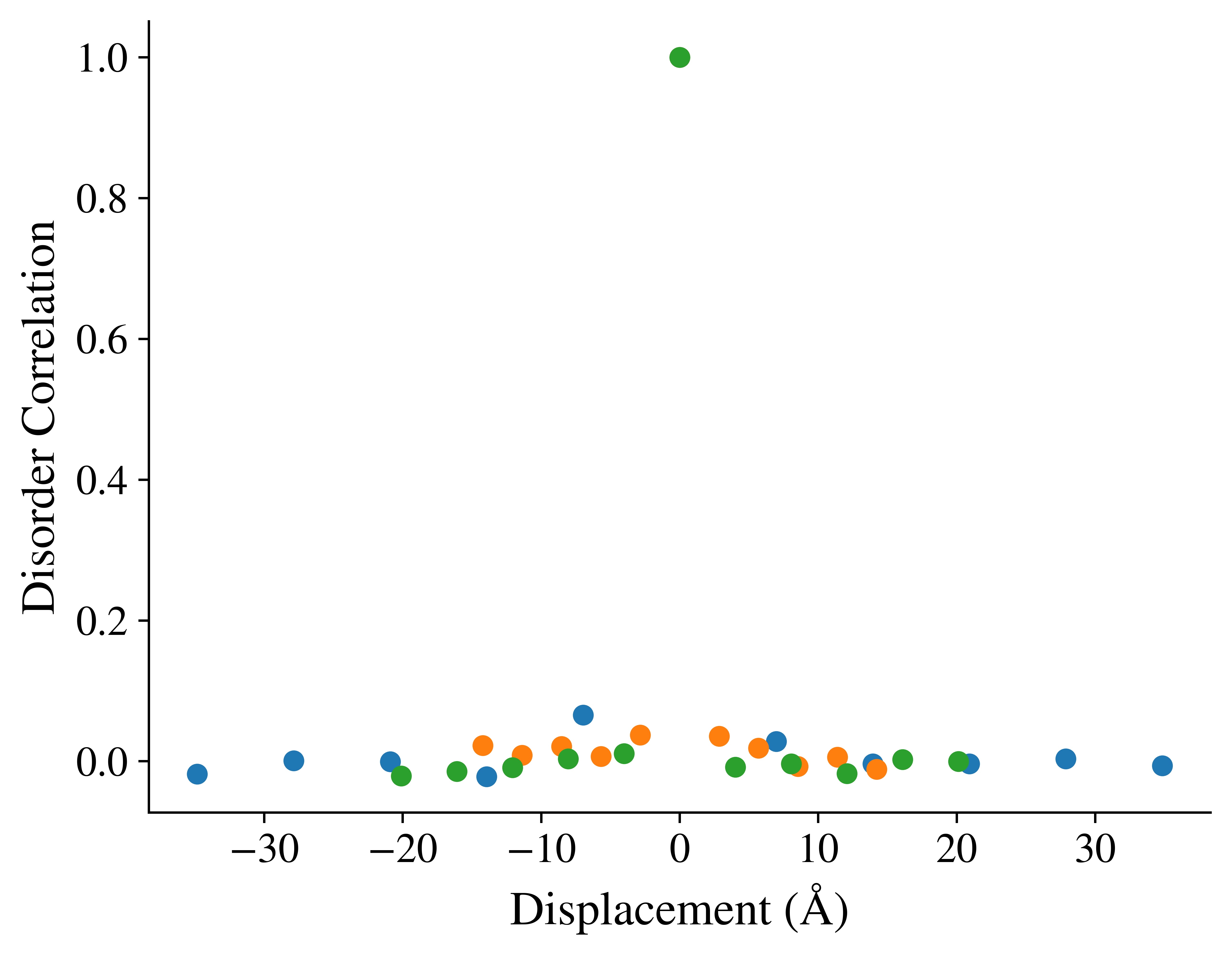}
  
  \caption{Numerical estimates of the disordered field winding correlations at $t=5\mu$s and $t=75\mu$s. The correlations are taken along three different traces through the crystal: along [111] (blue), [1$\bar{1}$0] (orange), and [100] (green).}
  \label{fig.sin_disorder}
\end{figure*}

For the disordered state technique, however, we require stronger conditions on the structure of disordered field correlations. Namely, we want to ensure that the disordered field phase-tagging is truly uncorrelated, even if the underlying disordered field is $r^{-3}$ correlated. As discussed in the previous section, we hope to enforce that at long enough times $t$,
\begin{equation}
    \mathbb{E}\bigg[\sin \omega_i t\sin \omega_jt\bigg] \approx \delta_{ij}.
\end{equation}
Due to the intrinsic correlations, this certainly is not true at short times, when $t \ll 1$. We can estimate this sinusoidal correlation function with the same Monte-Carlo methodology used to generate the distribution itself. In Figure \ref{fig.sin_disorder}, we verify that significant correlations are present at $t=5\mu$s of winding, and demonstrate that these correlations almost entirely vanish at $t=75\mu$s. Thus, we have strong numerical evidence that the disordered state technique is able to properly cancel correlations and probe directly the local auto-correlation signal.

\section{Hamiltonian Engineering Fidelity}
With the disordered field adequately characterized both numerically and experimentally, we turn our attention to careful calibration of Wei16. This sequence has been used extensively in the past on fluorapatite \cite{Wei2018, peng2019comparing, Peng2023, Stasiuk2023, Stasiuk2024}, which has slightly weaker interactions and fewer neighbors per-site such that Hamiltonian engineering errors were not a main concern. In the lithium fluoride system, however, we have found that Trotter and pulse error can be significant without proper mitigation strategies. For these reasons we have needed to lengthen our pulse duration to 1.1 $\mu$s, and reduce the delay primitive from $5\mu$s to $4\mu$s, as described previously in Section \ref{sec:ham_control}. Since the goal of this study is to explore regions of strong disorder, we also want to engineer very small interaction strengths -- as low as 1-2\% of the intrinsic interaction strength. Varying these parameters requires careful control of interpulse delays, which have a resolution of only $12.5$ns stemming from the onboard 80 MHz signal generator. Some basic F-matrix analysis, including rounding of all delays to the nearest 12.5 ns, shows that we should expect some small errors in the disorder-to-interaction ratio, on the order of 1-5\%. This analysis assumes perfectly square pulses, perfect tune-up, and only minor timing errors -- it would be best to calibrate experimental errors experimentally.

Direct calibration of the rescaled interaction strength in the Wei16 average Hamiltonian is not precise -- measurement of $\avg{\hat{S}_x(t)\hat{S}_x}$ under dipolar interactions reproduced the FID signal, the fundamental signal in classical NMR studies, but has no closed form representation for fitting. The approximate representation, a sinc function modified by a Gaussian, can be associated with statistical moments of the interaction Hamiltonian; such a description does not suit our needs well \cite{abragam1961principles}. Instead, we treat the details of the line-shape agnostically and focus on the collapse of the time/energy rescaling of the measured signals. Namely, if $\bar{\mathcal{H}} = c_2\mathcal{H}$, then all experimental curves should coincide when plotted against $c_2 t$. We distinguish these timescales as ``clock time'' (when referring to $t$) and ``virtual time'' (when referring to a rescaling of $t$ by the energy rescaling).

\begin{figure*}[t]
    \centering
    \includegraphics[width=0.32\textwidth]{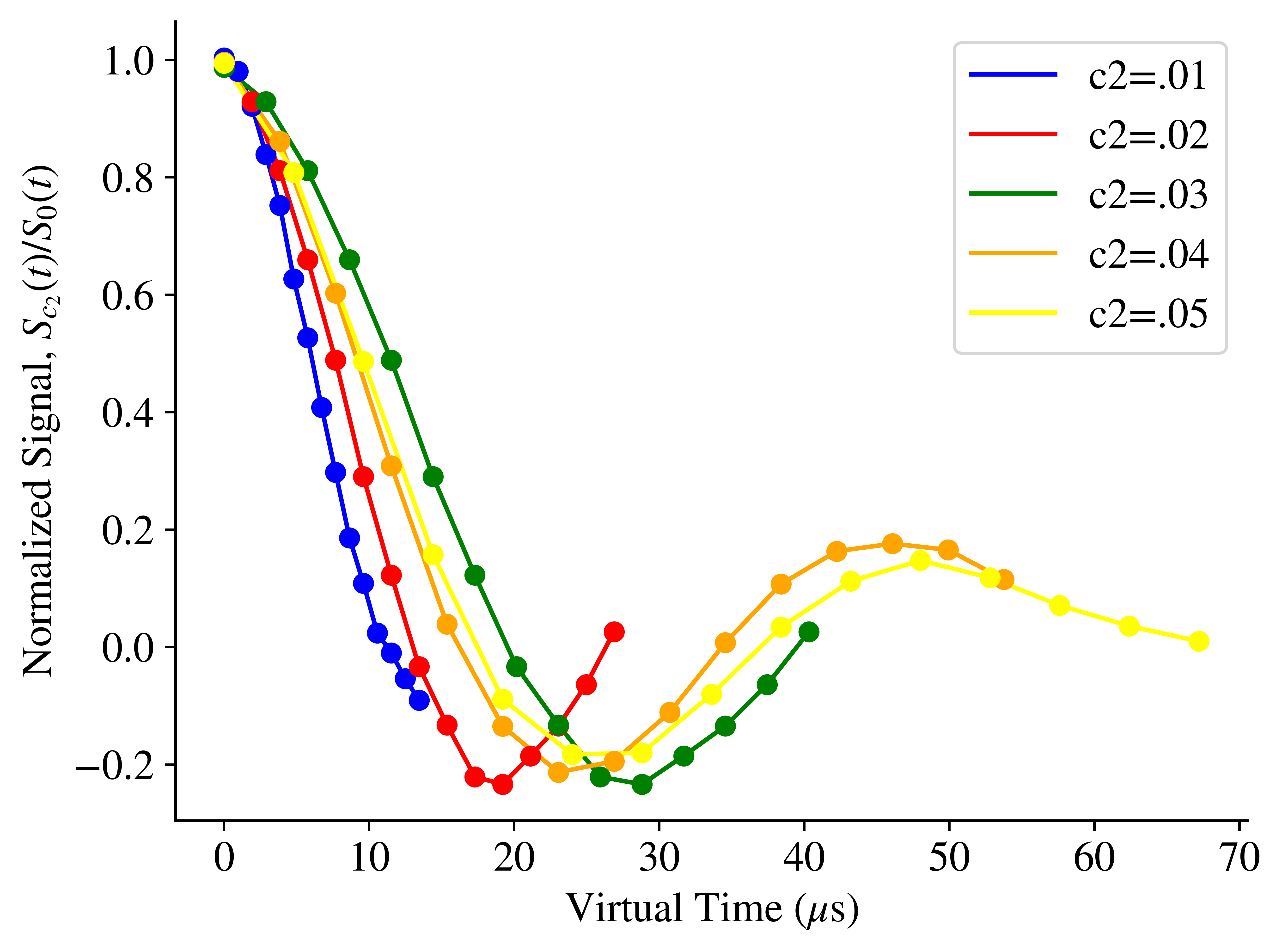}
    \hfill
    \includegraphics[width=0.32\textwidth]{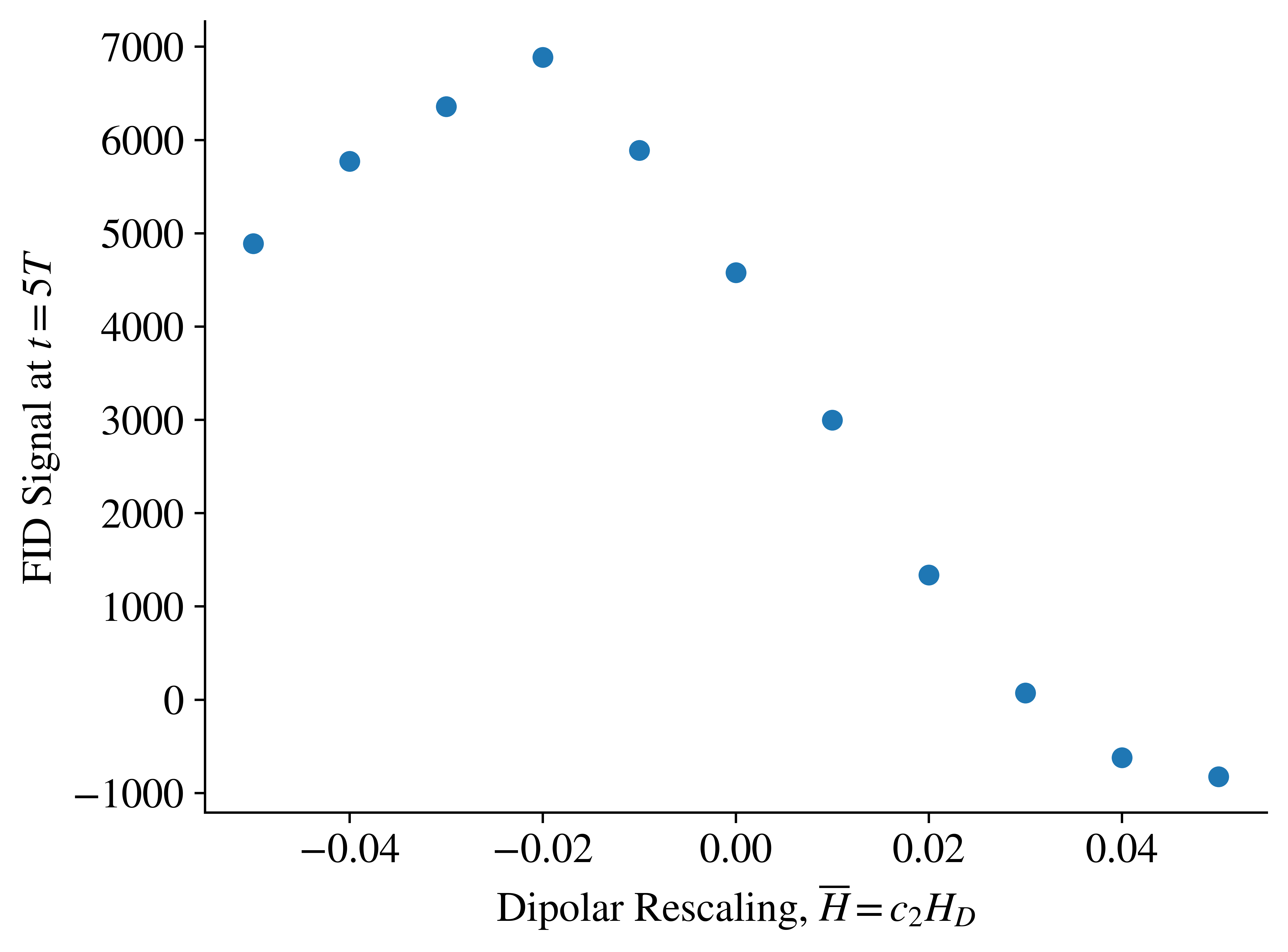}
    \hfill
    \includegraphics[width=0.32\textwidth]{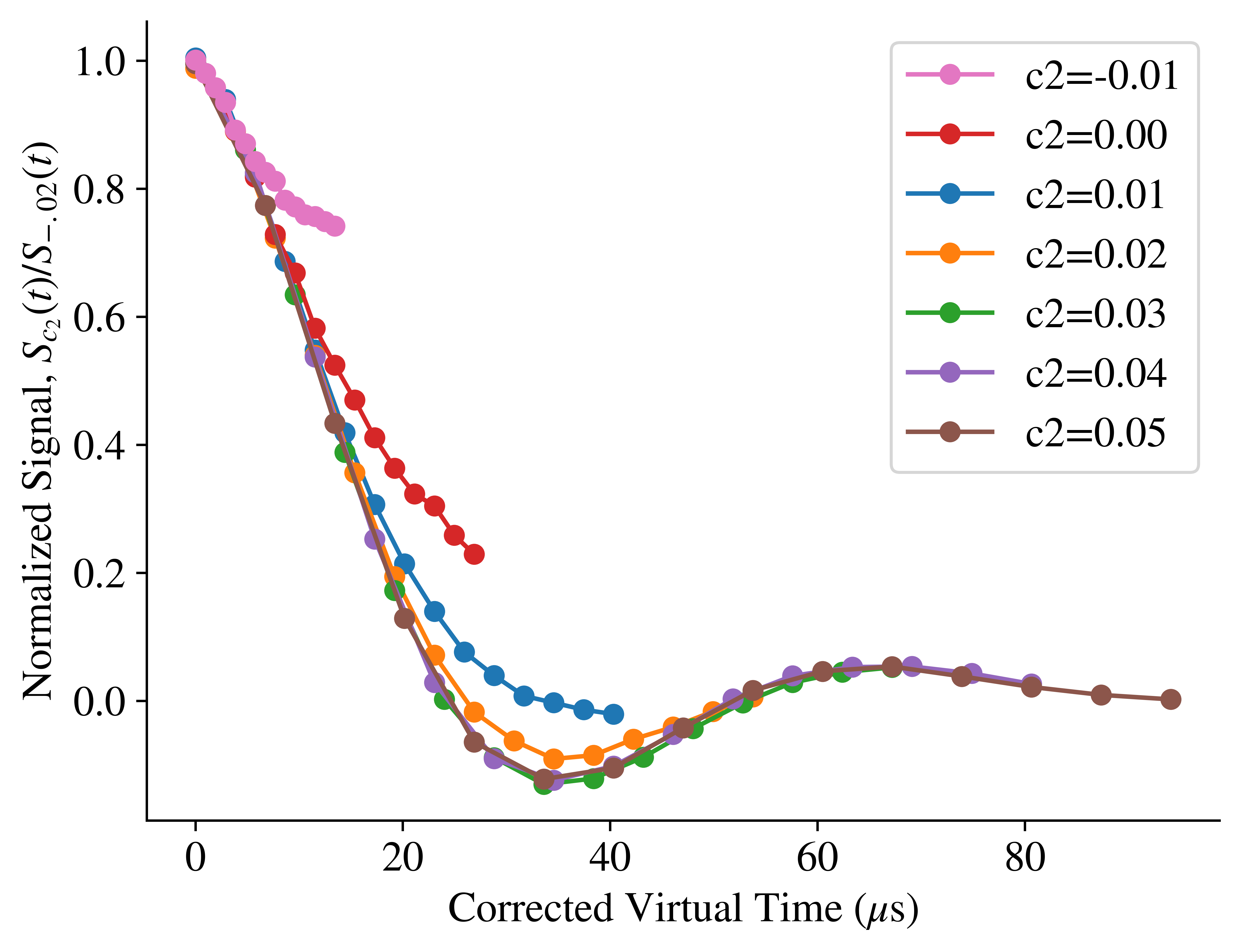}
    
    \caption{Left: The free induction decay (FID) signal of the sample under only the dipolar interaction, achieved with the Wei16 engineering sequence. Each curve represents a different overall rescaling of the interaction strength, and is normalized and rescaled in attempt to collapse each trace onto a single curve. Center: The FID signal at $t=5T$ (5 Floquet periods of evolution), as a function of the overall energy rescaling. If no offset error exists, this curve should be maximal at $c_2=0$. However, we see that the maximum occurs when $c_2=-0.02$, indicating there is an offset error in the overall rescaling of the average Hamiltonian. Right: A reproduction of the plot in the panel on the left, albeit with the adjusted energy scale offset and proper suspension signal normalization. These data collapse much better onto a single curve, as expected for a high-fidelity quantum simulation.}
    \label{fig:fid_test}
\end{figure*}

While canceling all disorder ($c_1=0$), we vary $c_2\in \lbrace0,.01,.02,.03,.04,.05 \rbrace$, and measure the FID signal $\avg{\hat{S}_x(t)\hat{S}_x}$. We plot these results directly as a function of clock time, as well as rescaled in both time and magnitude. Namely, we normalize the signal by what we believe is the decoupling signal, which should further normalize the error per Floquet period. The results of these experiments are shown in the left panel of Figure \ref{fig:fid_test}, demonstrating clearly that there are some significant errors - the experimental curves do not collapse under the \textit{a priori} correct rescaling. Visually, it seems that the smaller interaction strengths see the largest error, leading us to believe there is a finite offset in the applied dipolar interaction strength, $c_2 \longrightarrow c_2 + \epsilon$. To calibrate this offset error, we fix our measurement of the FID signal to $t=5T$, where $T$ is the length of a Floquet period, and vary $c_2$. We expect that when $\bar{H} \approx 0$, the resulting signal should be maximal. This point in time was chosen as a balance between distinguishability between interaction strengths and total signal magnitude. Panel (b) of Figure \ref{fig:fid_test} shows a maximum value occurs at $c_2\approx -0.02$, indicating that $\epsilon = .02$. Finally, we verify the validity of this correction technique by repeating the full FID experiments for $c_2\in\lbrace -0.01, -0.02\rbrace$, using the $c_2=-0.02$ data to normalize the FID curves, and adding in the $\epsilon = 0.02$ offset in the time rescaling. These results are shown in panel (c) for Figure \ref{fig:fid_test}, and show a much better collapse profile, indicating that our Hamiltonian engineering has  high fidelity up to some calibrate-able offsets. For traces where the interactions are nearly entirely decoupled, we attribute the deviations with increased pulse error accumulated during the course of evolution. Additionally, the $c_2=-.02$ data fits much better to a simple exponential decay than the $c_2=0$ data, with a decay timescale of 1323 $\mu$s compared to 528 $\mu$s at $c_2=0$. Attempts to perform a similar analysis for $c_1\neq0$ are plagued by normalization rescaling issues, similar to the behavior in the StaBerYXX calibration experiment, and suffers from an additional $D_y$ error term as described in Table \ref{tab:first-order}. We thus take the $c_2$ shift calibrated at $c_1=0$ for all disorder strengths.

\section{Random Walk Model and Percolation Numerics}\label{sec:random_walk}

Here, we provide details of the random walk model developed in the main-text and further describe our numerical investigations. To start, we consider a pair of identical spin-1/2 nuclei, ignoring all other spins for the moment. Within the rotating frame, non-trivial evolution only occurs when the state of the spins are anti-aligned, e.g. in the manifold spanned by $\lbrace \ket{01}, \ket{10}\rbrace$. Namely, these excitations will exchange with each-other under a ``flip-flop'' process, mediated by the $\hat{S}_x^{(1)}\hat{S}_x^{(2)}+ \hat{S}_y^{(1)}\hat{S}_y^{(2)}$ term of the Hamiltonian. In the classical description of spin diffusion, the two states are smeared by a width $\overline{\Delta \omega}^2$ defined by the Van Vleck formula for the second moment \cite{VanVleck1948,abragam1961principles}. Using time-dependent perturbation theory, the two-isolated spins exchange excitations with a probabilistic rate $W$ over a lattice spacing $a$, and hence the diffusion coefficient is loosely proscribed by $D\approx Wa^2$, which is generally in the range of $10^{-13}$ to $10^{-12}$ cm$^2$/s \cite{abragam1961principles}. This model has been generally successful in qualitative and approximately quantitative predictions for relaxation in solid-state NMR. Significant extensions have been made to the model in order to better estimate $D$, as well as predict the diffusion of energy in addition to that of magnetization \cite{boutis2004spin, Greenbaum2005}. However, these classical studies crucially lack a tuneable disordered field knob; the only ``free'' parameters in this description are the orientation of the sample relative to the magnetic field and its internal crystal geometry. 

Thus, we extend the classical spin-diffusion model by first introducing a local disorder term in the restricted two-spin Hamiltonian,
\begin{equation}
    \mathcal{H}=(\omega_1S_z^1+\omega_2S_z^2)-\frac{J}{4}\left(\hat{S}_+^{(1)}\hat{S}_-^{(2)} +  \hat{S}_-^{(1)}\hat{S}_+^{(2)}\right) \cong \frac{1}{2}\begin{bmatrix}
    \delta\omega & -J/2 \\
    -J/2 & -\delta\omega
    \end{bmatrix},
\end{equation}
where $\delta\omega = \omega_1-\omega_2$, with matrix representation of the Hamiltonian defined on the anti-aligned manifold. Under the assumption of i.i.d. disorder, as expected for a disordered field generated by a secondary nuclear spin species, $\delta\omega \sim N(0,2\sigma^2)$ is a random quantity. As a restricted two-level system, the \textit{maximum} probability of exchange can be geometrically deduced on the Bloch-sphere, such that
\begin{equation}
    \mathbb{P}[01 \longrightarrow 10] = \sin^2\theta = \frac{1}{1+4(\delta\omega/J)^2}.
\end{equation}
The average probability of exchange is one-half of this quantity.

Now, to estimate the trend of the diffusion coefficient as a function of disorder, we first insist that these spins will exchange their excitation, and then determine how long this swap is expected to take. Namely, the maximum swap probability is $\sin^2\theta$, which is achieved after a time $\tau = \frac{2\pi}{J}\sin\theta$. At this time $\tau$, we insist that either the exchange has occurred or the state is reset and we repeat the process indefinitely. The probability that the swap happens at the $k$-th try is $\sin^2\theta \cos^{2k}\theta$ and we will have waited $t=k\frac{2\pi}{J}\sin\theta$. Hence, $k\sim \text{Geom}(\sin^2\theta)$ is a geometrically distributed random integer, with expectation value
\begin{equation}
    t_{swap} = \mathbb{E}[k\tau] = \frac{2\pi}{J\sin\theta}.
\end{equation}
Of course, $\sin\theta$ is itself still random due to the disordered field dependence. Under a disorder average, $t_{swap}$ takes the form of the Tricomi confluent hypergeometric function, depending only on the ratio $\frac{4\sigma}{J}$. Numerically, we find that for large enough disorder, the time to swap grows linearly in the disorder strength,
\begin{equation}
    \frac{J}{2\pi}\avg{\mathbb{E}[k\tau]}_{\omega} = 4\frac{\sigma}{J}\text{HyperGeomU}\left(-1/2, 0, \left( \frac{J}{4\sigma}\right)^2\right) \sim 2.25\frac{\sigma}{J}.
\end{equation}
By characterizing the mean time to swap along with the mean distance traveled during a swap, it is straightforward to directly estimate the diffusion coefficient, given as
\begin{equation}
    D = \frac{\avg{l^2}}{2\avg{t_{swap}}}.
\end{equation}
For simplicity, we compute the diffusion coefficient using a nearest neighbor approximation, $J=J_{nn}$, with average separation of $r = \frac{\sqrt{2}}{2}a$, where $a=4.02$ \AA. Thus, for sufficiently large disorder, we estimate $D \approx D_0/(\sigma/J)$, where $D_0 \approx 2\times10^{-12}$ cm$^2$/s is computed using the exact value of the hypergeometric function at 0 disorder. This value is inexact -- a more precise estimate should include the effects of homonuclear broadening as well as more distant hopping. Given that both of these additions compete with each-other, and the fact that the zero-disorder estimate has the correct order of magnitude \cite{Klempt2003}, we believe that this model qualitatively captures the behavior of the rate of spin diffusion in the presence of disorder.

Now, let's imagine the disordered field is very strong -- when the detuning between two sites is much larger than their coupling strength, the probability of an exchange occurring becomes quite small. In this picture we imagine two types of spin-pairs: those which can exchange excitations with each-other, and those that cannot. In this regime, we prescribe a recipe for determining the probability that a pair of spins are coupled strongly enough to swap excitations as a function of the disorder strength. Direct averaging of the maximal swap probability results in what we call the ``bond acceptance probability'',
\begin{equation}
    p_b(\sigma,\bm{r}) = \avg{p_{max}(\delta\omega)}_\omega = \frac{J(\bm{r})\sqrt{\pi}}{4\sigma}e^{\frac{J(\bm{r})^2}{16\sigma^2}}\text{Erfc}\left( \frac{J(\bm{r})}{4\sigma}\right).
\end{equation}
This description allows us to consider arbitrary connectivity and disorder in a consistent and uniform manner. Given the bond acceptance probability $p_b(\sigma,\bm{r})$, we can now numerically determine at what disorder strength the system will percolate.

In an infinitely large regular lattice, if bonds are removed uniformly and independently with some probability, then we expect that at some critical value the system will shatter into independent disconnected components. By mapping the effect of the disorder field and bond-geometry into a single probability $p_b$, the emergent hydrodynamics of the random walk depends intimately on whether the system is percolated. If no infinite cluster exists, then every random walker on the disordered lattice can only reach a finite distance from their starting point, hence the dynamics are localized. Otherwise, an infinite subgraph exists, such that walkers on the infinite subgraph will behave diffusively at sufficiently late times. Hence, the onset of percolation, that is, the shattering of the infinite lattice, should determine the disorder strength at which localization occurs (be it Anderson, Stark or Many-Body Localization). 

Numerically, we generate a graph of fixed connectivity, where each vertex contains its spatial location, and each edge contains the non-dimensional coupling strength of the two vertices it connects. This data payload structure is enabled by the \verb|rustworkx| graph library, which is efficiently written in Rust, and allowed our graphs to easily scale up to 17,969 vertices \cite{Treinish2022}. We vary the base graph size in order to reduce our sensitivity to finite size effects, as percolation only formally occurs on an infinite lattice. More efficiently written algorithms and parallelization could further increase the graph size by a few orders of magnitude, though such intense scaling is not needed for this analysis. Then, over 100 independent trials, we randomly cut bonds with probability $1-p_b(\sigma,\bm{r})$, and partition the resulting graph into disjoint subgraphs $G = \bigcup_i G_i$. The size and frequency of this partition fully determine the percolation state functions, the percolation strength and the average cluster size.

\begin{figure}[t]
    \centering
    \includegraphics[width=0.56\linewidth]{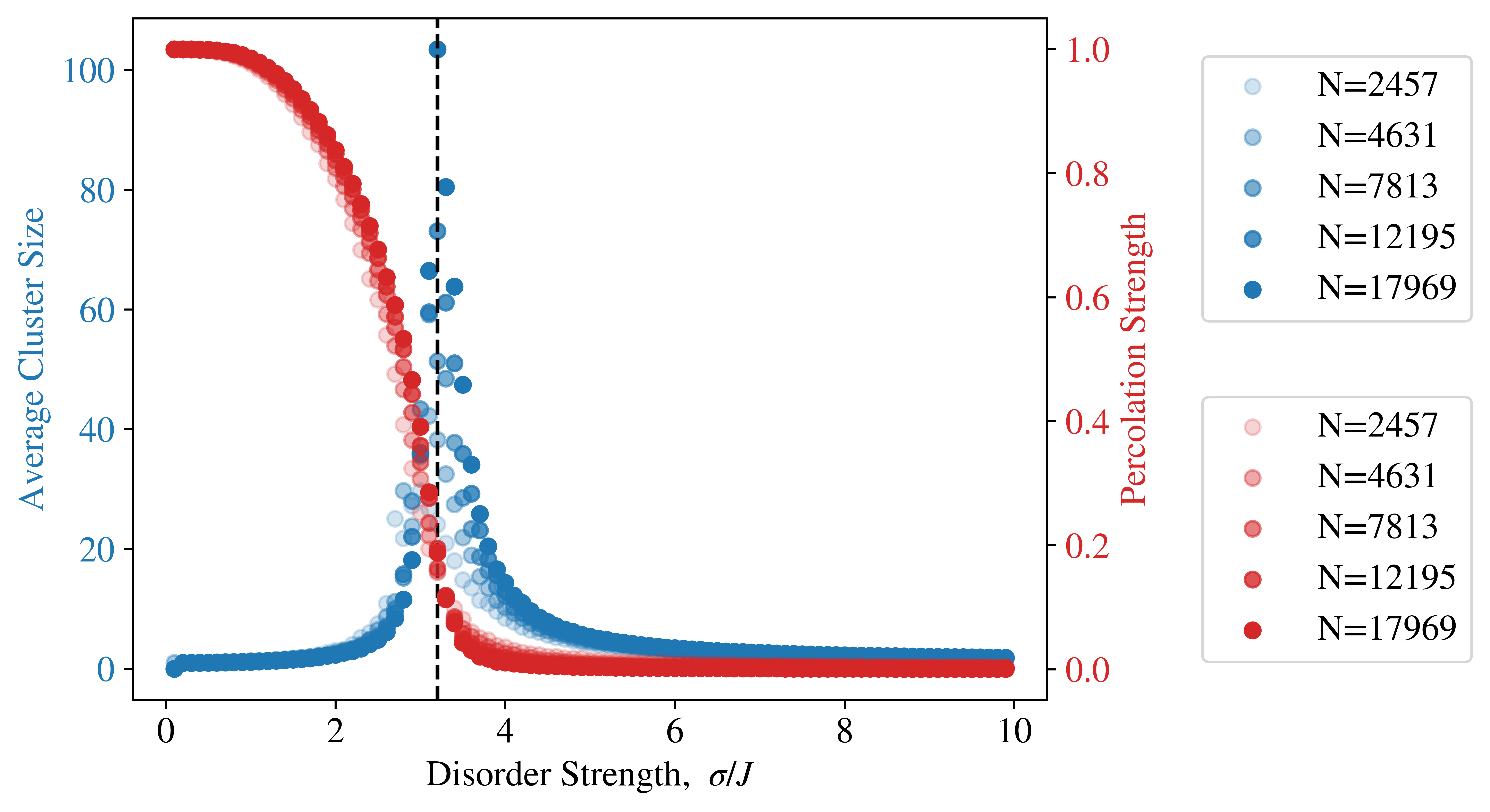}
    \hfill
    \includegraphics[width=0.4\textwidth]{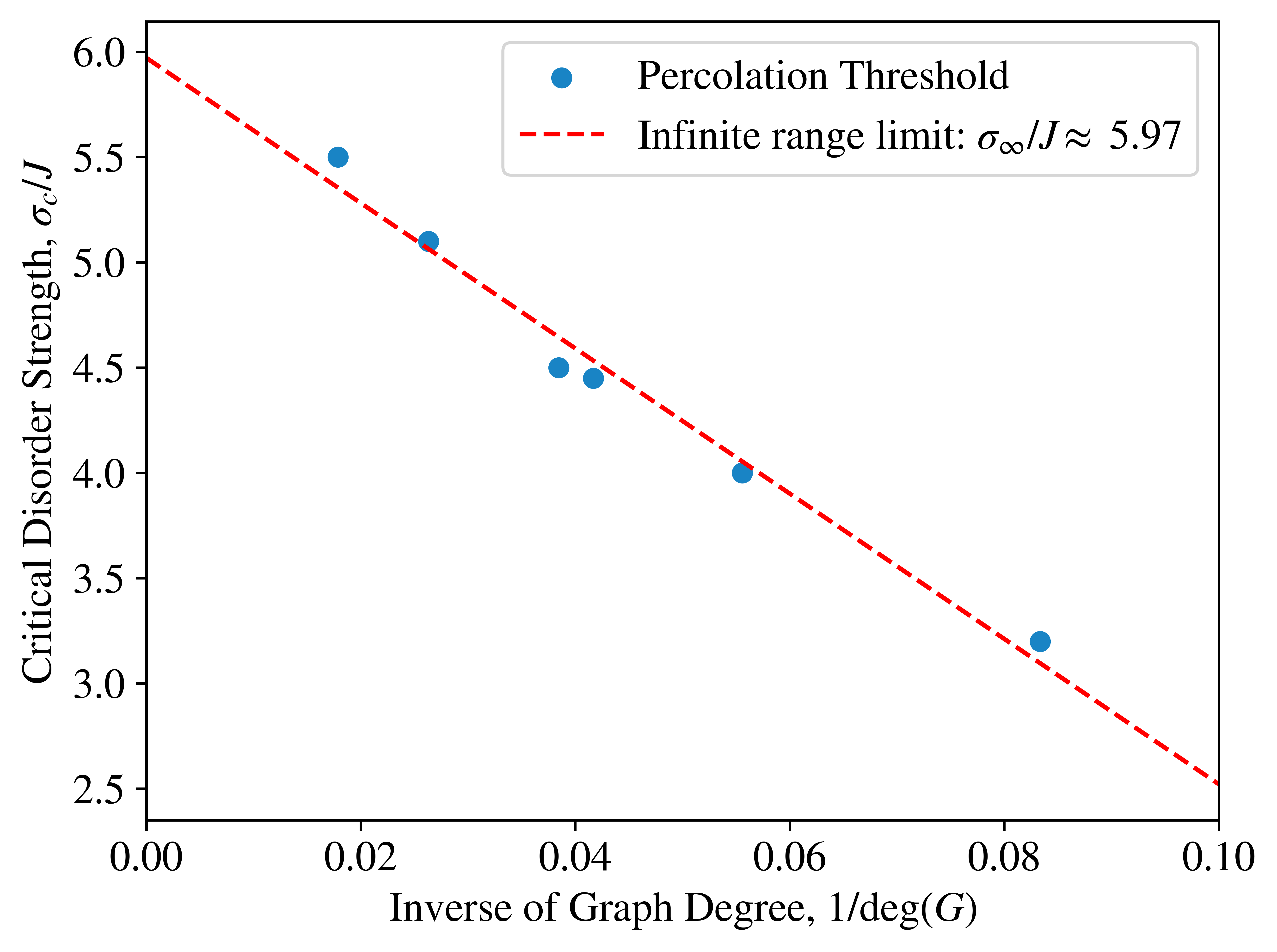}
    \hfill
    \caption{Left: Percolation phase transition for a nearest neighbor FCC spin network, using bond acceptance probability $p_b(\sigma)$. The critical point, $p_c = p_b(\sigma_c/J)$ occurs at $\sigma_c/J \approx 3.2$. Right: The critical point as a function of bond topology, specifically the inverse of the number of edges of an interior site (graph degree). The data appears to follow a linear trend; interpolation of this curve predicts that the infinite-range $1/r^3$ connectivity limit has a percolation phase transition at $\sigma_\infty/J \approx 5.97$.}
    \label{fig:nearest_neighbor_perc}
\end{figure}

The critical point of the  percolation phase transition can be found by estimating the average cluster size $S_N$. We numerically compute $S_N$ by counting the size of disconnected spin networks in the generated graphs $G$. Given the disjoint union, $G = \bigcup_i G_i$, we can compute the size of each connected cluster $s_i=|G_i|$. By estimating the cluster size distribution probability $n_s$ from numerical histograms over many instances of the graph $G$ at a given bond acceptance probability, we compute finite size approximations of the percolation strength (the probability that a site belongs to the ``infinite'' percolating cluster) $P_N(\avg{p_b})$, and the average (finite) cluster size, $S_N(\avg{p_b})$ \cite{christensen2002percolation}:
\begin{align}
    P_N(\avg{p_b}) &= \frac{\text{max} \abs{G_i}}{\abs{G}} = \frac{\text{max} \abs{G_i}}{N}\\
    S_N(\avg{p_b}) &= \bigg(\sum_{s=1}^\infty s^2 n_s(\avg{p_b})\bigg) \bigg/ \sum_{s=1}^\infty s n_s(\avg{p_b})
\end{align}  
To avoid counting the finite size analog of an infinite cluster, the largest cluster size is excluded in the counting statistics of each trial. In the limit $N\to\infty$, $P(\avg{p_b}) = 0$ for $\avg{p_b} < p_c$, and $S(p_c)\to\infty$. Since the average non-infinite cluster size diverges at the critical point, the critical disorder strength is estimated as the disorder strength which maximizes $S_N$. At or near this point, $P_N$ becomes scale-invariant, a usual hallmark of the emergence of a phase transition, and provides additional evidence of the emergence of a phase transition in the graph structure. Characterization of these state functions fully defines the percolation phase transition. Importantly we are able to extract the location of the phase transition with high confidence. By considering larger system sizes and focusing on a narrow range of disorder strengths about the phase transition, we could also extract the critical exponents of the phase transition, though these information is outside the scope of the current study. 

In principle, the finite size analog of the infinite subgraph should be connected to each face of the boundary, though this is not always the case due to finite size fluctuations. By always discarding the largest subgraph, we avoid making this determination at the cost of discarding more information than necessary, introducing some small systemic error into our analysis. We do not expect this to meaningfully impact the location of the phase transition, at least with respect to the resolution probed here. The expected behavior of these functions in the limit $N\to\infty$, is that $P(p_b) = 0$ for $p_b < p_c$, and $S(p_c)\to\infty$. Since the average non-infinite cluster size diverges at the critical point, the critical disorder strength is estimated as the disorder strength which maximizes $S_N$. The nearest neighbor percolation data is shown in the left panel of Figure \ref{fig:nearest_neighbor_perc}. Then, this whole procedure is repeated another five times, increasing the number of initial bonds at each step, such that at the sixth nearest neighbor level, the degree of the graph is 56. ``Nearest'' is determined via bond-strength ordinality, such that nearest neighbor considers only the most strongly coupled sites, of which there are 12, and coincide with geometric nearest neighbors due to the [111] B-field orientation. Next-nearest neighbor coupling increases this by an additional 6 bonds per site, and which clearly do not coincide with all geometrically next-nearest neighboring sites. From this data, we interpolate and predict the infinite range limit, $\sigma_\infty/J \approx 5.97$, the details of which are shown in panel (b) of \ref{fig:nearest_neighbor_perc}.

\begin{figure}
    \centering
    \includegraphics[width=0.56\linewidth]{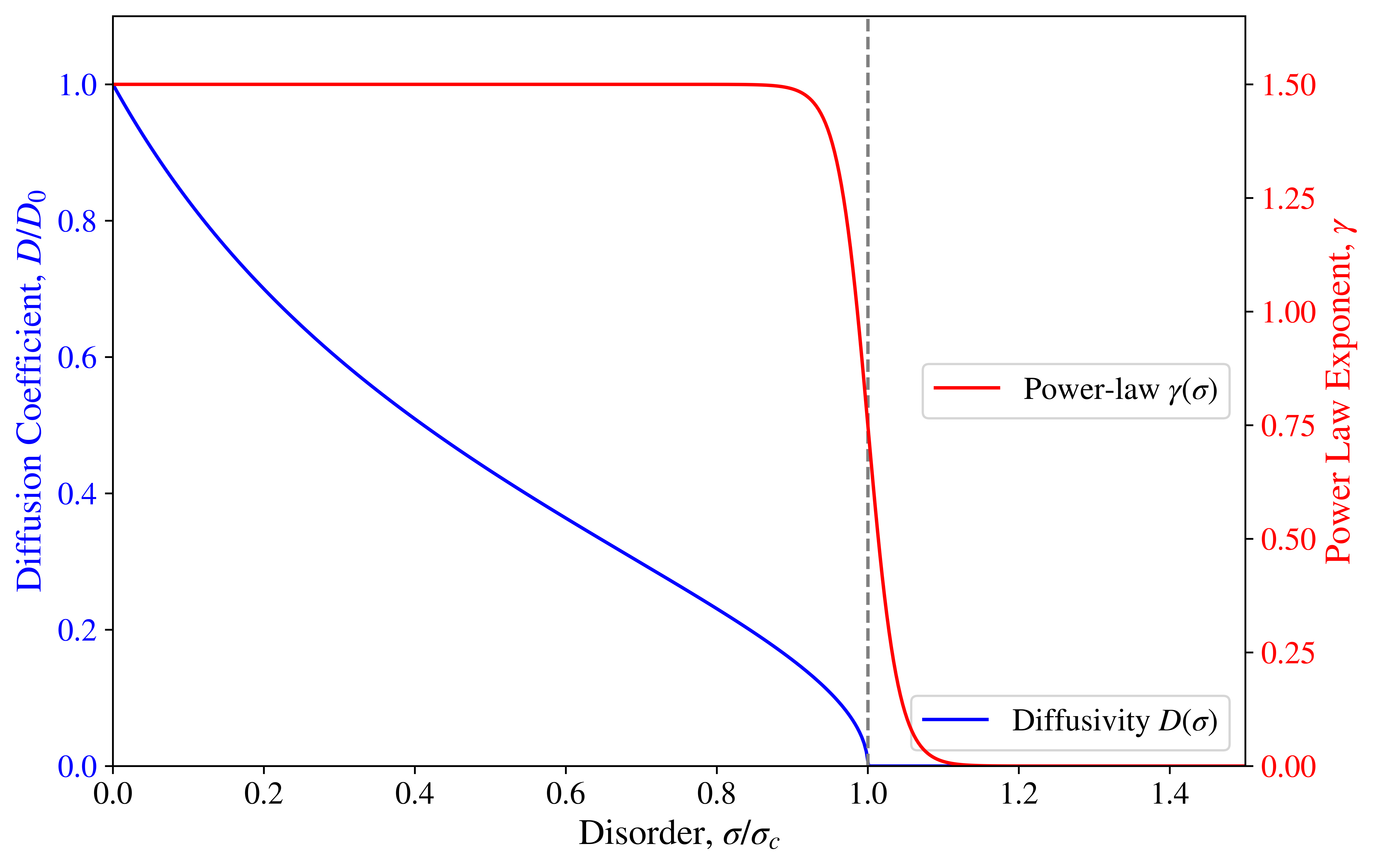}
    \caption{A sketch of the predictions on the transport properties of random walkers in a disordered dipolar spin system. At weak disorder, motion is diffusive, with $D(\sigma) \sim D_0/\sigma$. At sufficiently strong disorder, the system passes through a percolation phase transition, resulting in emergent localization}
    \label{fig:theory_inf_time}
\end{figure}

To summarize the model thus far, for weak disorder transport is expected to be diffusive, with a diffusion coefficient that slows as the inverse of the disorder strength. Then, at sufficiently strong disorder, the disordered field shatters the connectivity of the lattice, localizing motion. Relevant to the experiments of the main text, this should impact the survival probability via the effective diffusivity $D(\sigma)$ and the power law exponent $\gamma(\sigma)$. Concretely, the survival probability can be written in the following form, 
\begin{equation}
    S_d(t) \propto \left(\frac{1}{4 \pi D(\sigma) t}\right)^{\gamma(\sigma)}, 
\end{equation}
defined up to some constant of proportionality which fixes the units. It is precisely this undetermined constant of proportionality that inhibits our ability to extract the diffusivity experimentally, as indicated in the main text. For an observer at sufficiently late times, our predictions for $D$ and $\gamma$ are depicted graphically in Figure \ref{fig:theory_inf_time}.

In the main text, our experimental data shows that diffusion and localization are not the only transport possibilities. Namely, there seems to exist an intermediate regime of mobility, slower than diffusion, but certainly not localized. This type of motion is dubbed anomalous diffusion, specifically subdiffusive since $\gamma_{obs} < 3/2$. For random walkers on a disordered lattice anomalous diffusion is possible; diffusion is known to be anomalous for arbitrarily late times when the system is fine-tuned to the percolation point. This is because the infinite percolating network has a fractal structure which spans the entire extent of the system; random walkers on a fractal are known to be subdiffusive relative to a Euclidean lattice embedded in the same dimensional space \cite{Alexander1982}. For finite time observations, subdiffusive behavior can be expected over a timescale commensurate with the correlation length of the percolation phase transition.

\begin{figure}[t]
    \centering
    \hfill
    \includegraphics[width=0.48\linewidth]{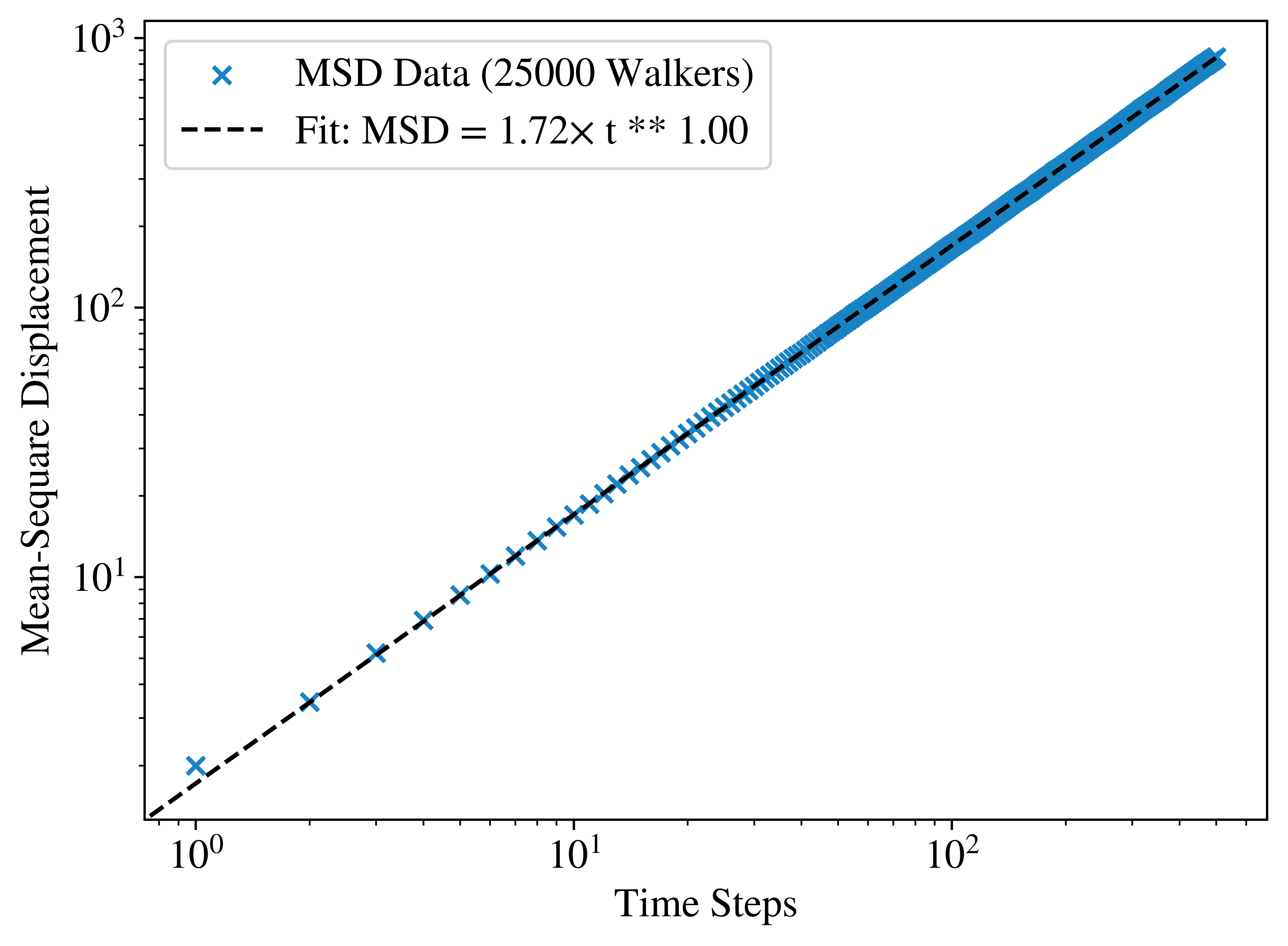}
    \hfill
    \includegraphics[width=0.48\linewidth]{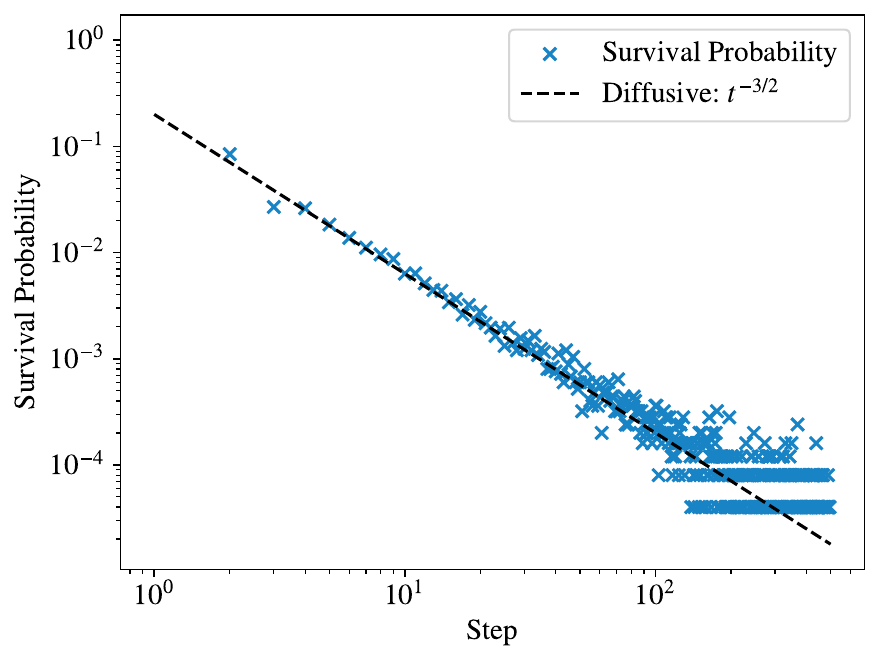}
    \hfill
    \caption{The estimated mean-squared displacement (MSD) and the survival probability for random walkers on an FCC lattice, with $N_{walker} = 25,000$ averages. In the left panel, we confirm with a power-law fit that the MSD grows linearly as a function of time, indicating diffusive behavior. In panel the right panel, we show that the survival probability is consistent with the expected diffusion autocorrelation of $t^{-3/2}$. Notice, however, that the autocorrelation signal decays to zero and the minimum resolution accessible via this procedure is $1/N_{walker} = 4\times 10^{-5}$. Hence, the survival probability is unfavorable for numerical extraction of the power-law $\gamma$ for arbitrarily late times, since it requires a $N_{walker}$ to grow polynomially with the final desired simulation time.}
    \label{fig:dense_rw}
\end{figure}

We investigate the finite time subdiffusive behavior with numerical simulations of a classical random walk on a percolating graph. In contrast to the percolation phase transition determination, we do not generate a dense lattice and then remove bonds, as this is computationally inefficient and has strong finite size effects even for graphs with more than 50,000 vertices. To verify the validity of this numerical scheme by checking that the resultant random walks average to diffusive motion, we investigate an ensemble of $N_{walker}=25,000$ random walkers on a fully connected and unbounded FCC lattice. Both in the mean-square displacement and the survival probability, our data is consistent with the diffusive motion. In Figure \ref{fig:dense_rw}, it is clear that both the MSD and the survival probability follow diffusive statistics: the MSD grows linearly in time, while the survival probability decays as $t^{-3/2}$. Because random walks are not recurrent on the 3D lattice \cite{lawler2010random}, collecting sufficient occurrences of returns to the origin for estimating the survival probability requires a very large number of walkers. Numerically, this makes the survival probability a relatively poor estimator of the power law $t^{-\gamma}$, requiring many more averages than the MSD to achieve similar confidence levels, and has a minimum resolution of $1/N_{walker}$. This can be seen directly in the spread of data at late times in the right panel of Figure \ref{fig:dense_rw}, which is visibly much larger than the MSD curve in left panel with the same number of random walkers. Experimentally, this resolution problem is not a concern until the signal-to-noise ratio reaches a value of around 1. Inherent in the NMR observable is a system-wide spatiotemporal average with over $10^{23}$ spins -- the statistics are not limited by the number of experimentally participating random walkers.

\begin{figure}
    \centering
    \includegraphics[width=0.48\linewidth]{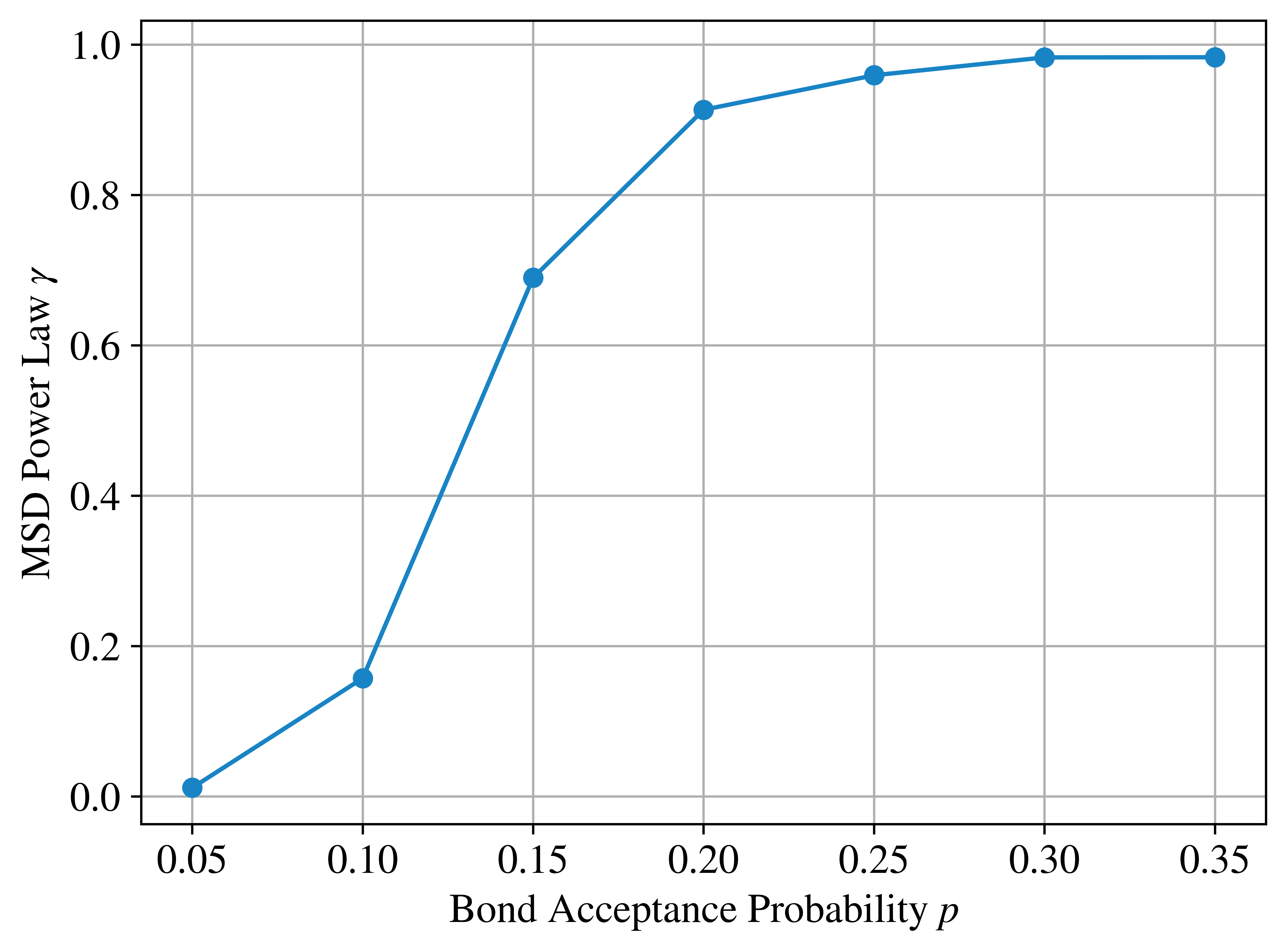}
    \hfill
    \includegraphics[width=0.48\linewidth]{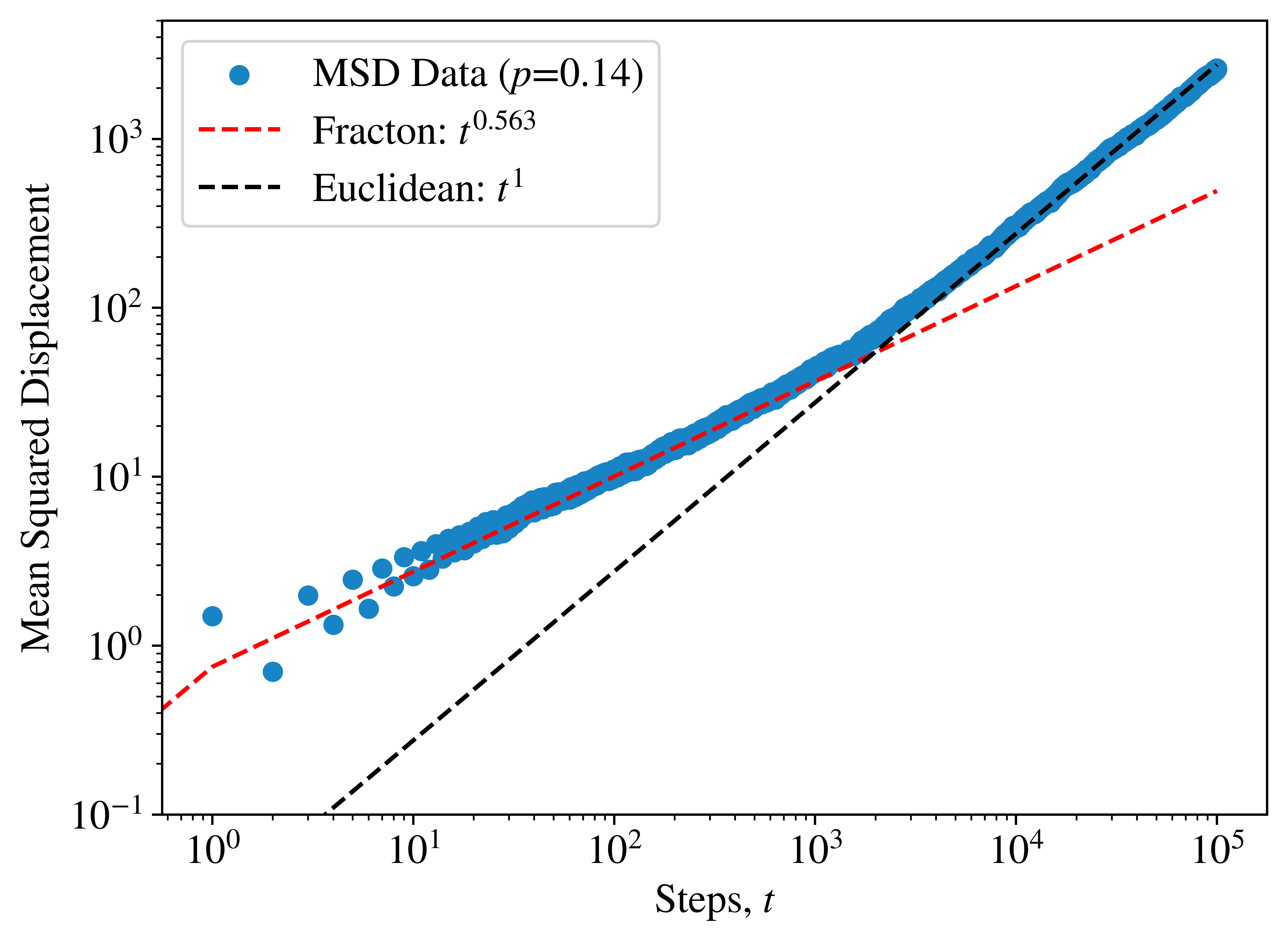}
    \hfill
    \caption{Investigations of the MSD power-law growth for variable disorder. To generate the left panel, we vary the bond acceptance probability and simulate 10,000 walkers for $t_f=1000$ time-steps each. The resulting MSD at each disorder strength is then fit to a single power-law, as shown. The data here indicates localization occurs between $p=.10$ and $p=.2$ and  consistent with the FCC percolation phase transition at $p_c\approx0.13$. For a better understanding of the apparent subdiffusive power-law, we also investigate the MSD of $N_{walker}=1000$ random walkers on a $p=.14$ disordered lattice up to time $t_f=100,000$. In the right panel the data clearly shows that a single power-law is a bad fit near the phase transition. There is subdiffusive regime at early times, persisting for about three decades, which then transitions into a diffusive regime for the final decade considered.}
    \label{fig:msd_disorder}
\end{figure}

Now, with diffusion at no disorder verified, we are ready to induce disorder into the lattice by random bond removal. As mentioned, it would be inefficient to generate an arbitrarily large disordered lattice. Instead, we define an expanding memory structure which saves the local state of nearby bonds. Each time a new site is visited, the locally connected bonds are probabilistically cut, and their state is stored. Of the remaining uncut bonds, the state chooses one at random and moves to that location. This procedure ensures that the lattice structure is quenched; cut bonds will not regenerate on a repeat visit. If, at the start, there are no valid bonds, we treat the system as fully localized. In Figure \ref{fig:msd_disorder}, we vary the bond acceptance probability and numerically simulate the mean squared displacement of random walkers on the disordered lattice, fitting the data to a single power law. The fitted power-laws are plotted as function of the bond acceptance probability in the leftmost panel of this figure, demonstrating clear evidence of a localization phase transition near the FCC percolation phase transition at $p_c\approx0.13$. In the right panel, we further investigate the apparent subdiffusive transport power law near the phase transition. Particularizing to a bond acceptance probability of $p=0.14$, we compute the MSD by averaging over 1000 walkers up to $t_f=10^5$ steps. The data clearly indicates two distinct regimes: subdiffusive at early times and diffusive at late times.

\begin{figure}[t]
    \centering
    \includegraphics[width=0.32\linewidth]{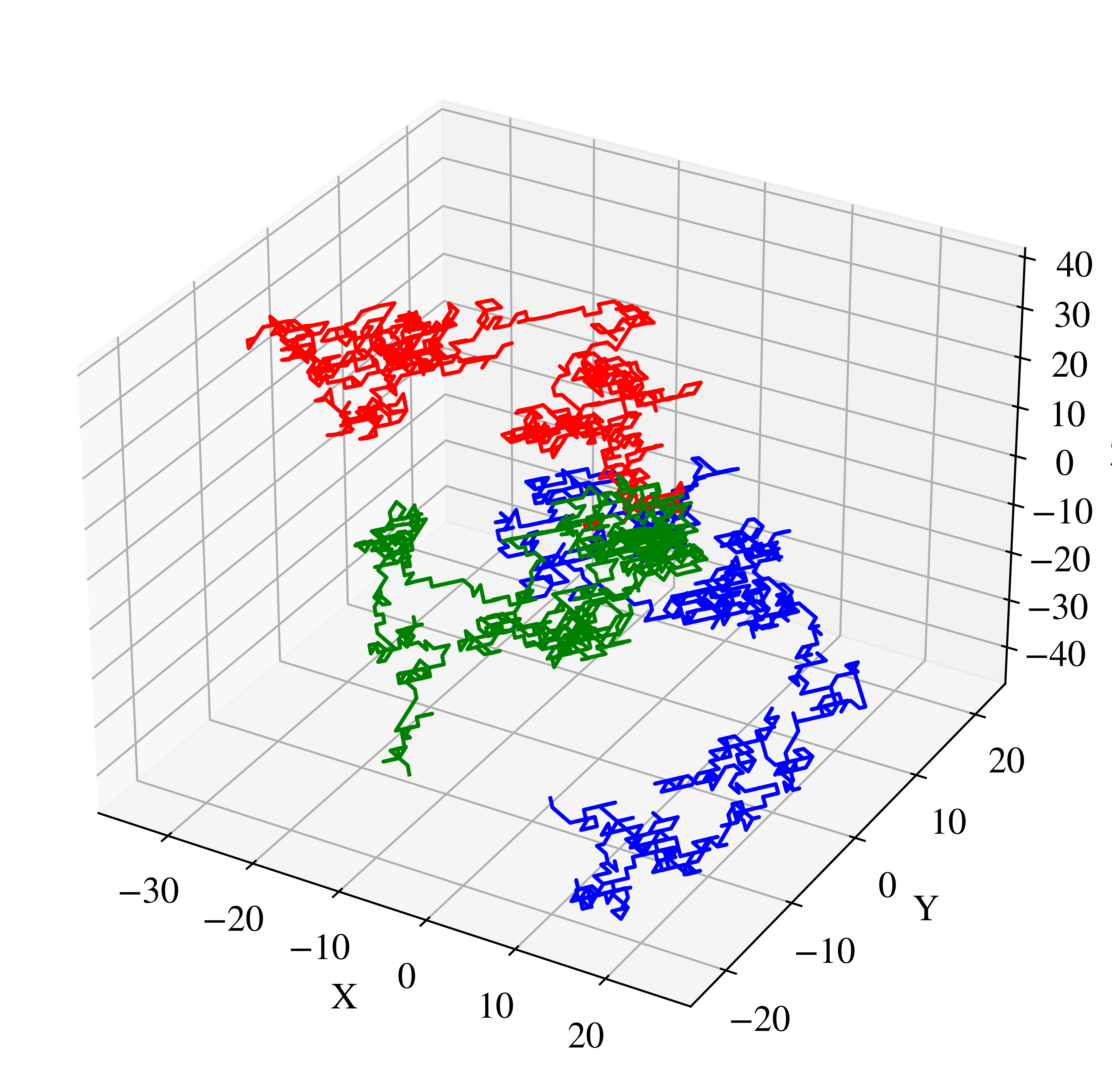}
    \hfill
    \includegraphics[width=0.32\linewidth]{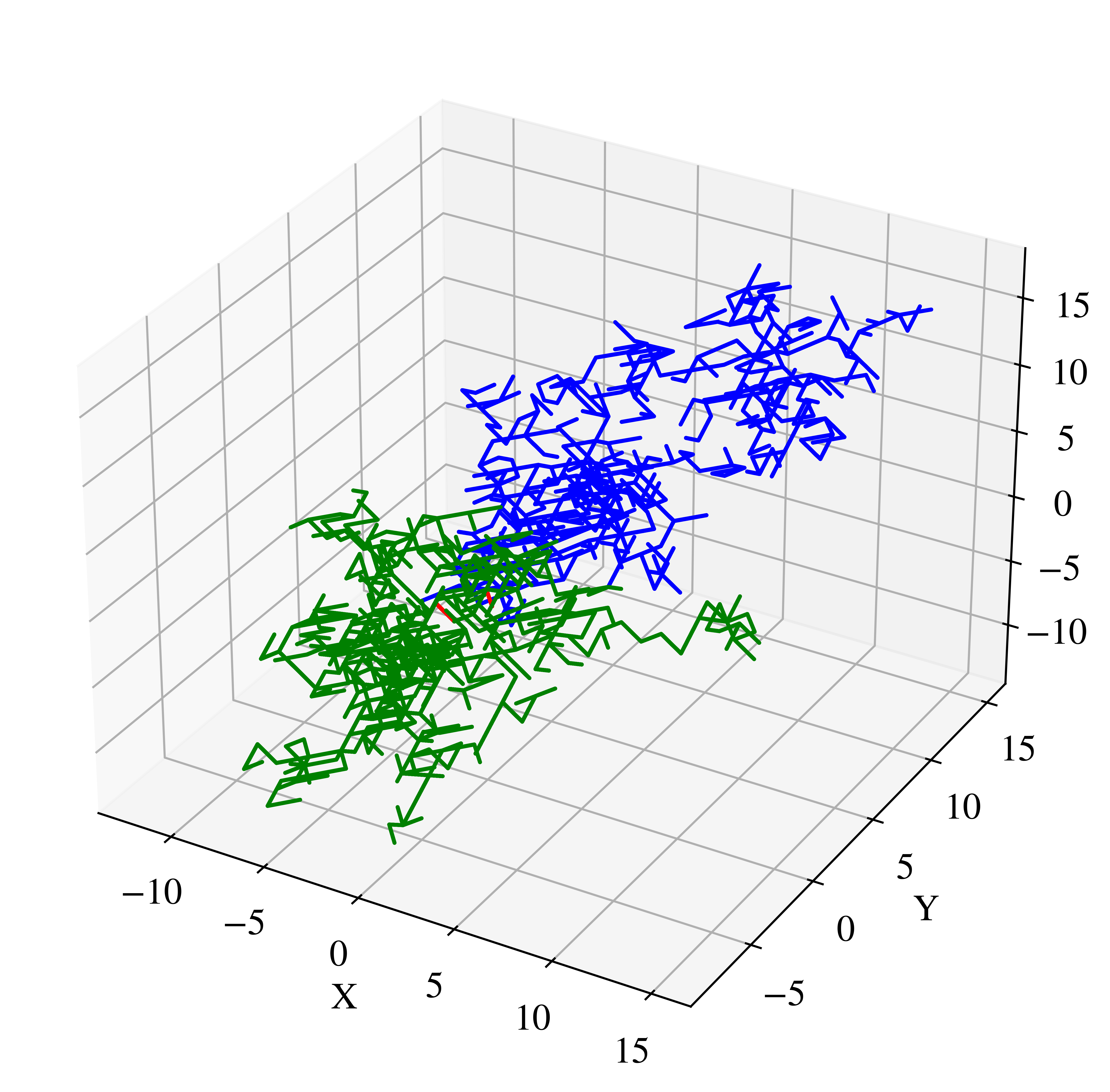}
    \hfill
    \includegraphics[width=0.32\linewidth]{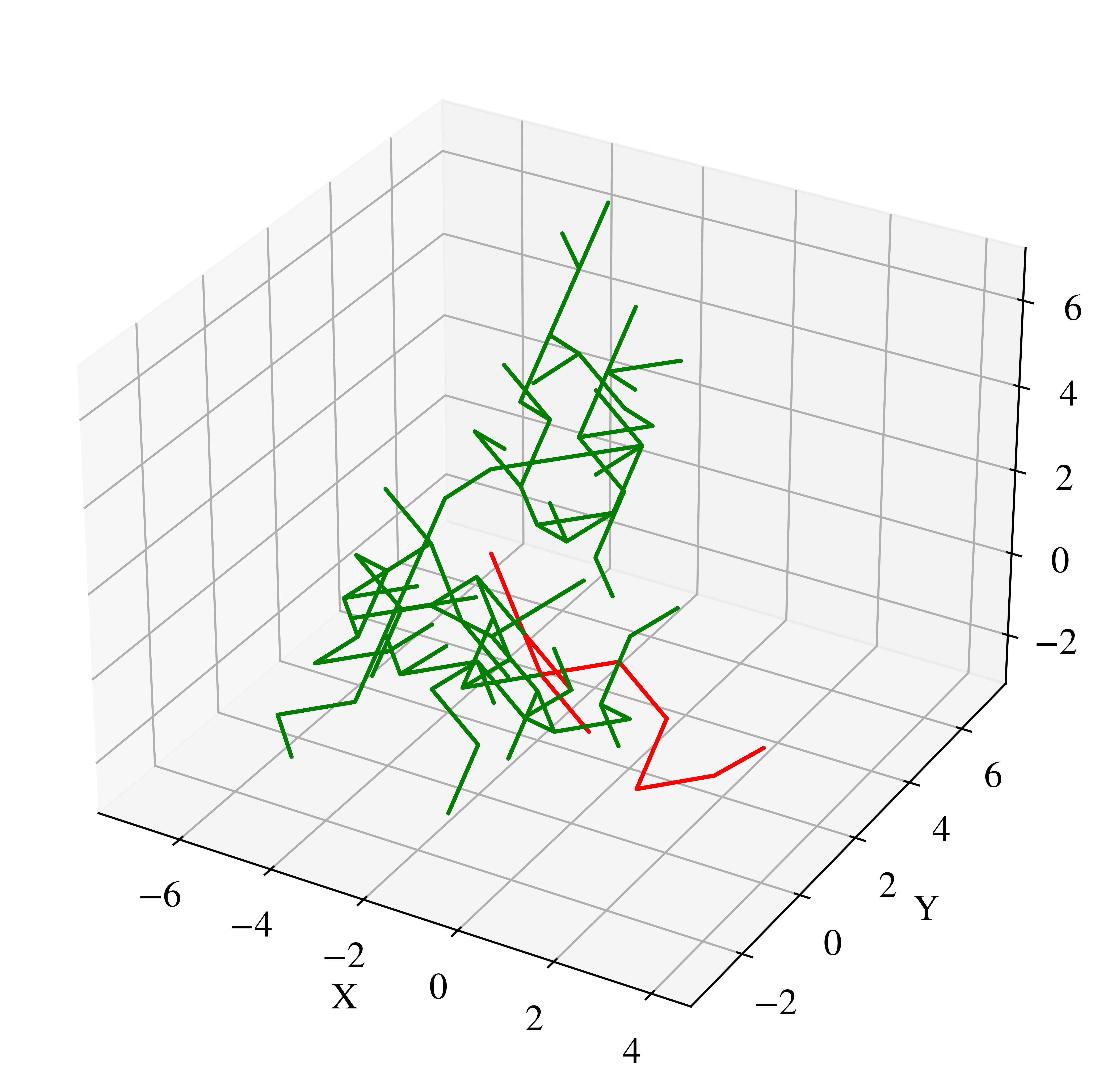}
    \caption{Random walk trajectories on the FCC lattice for different bond acceptance probabilities $p=$ 1 (left), 0.13 (middle), 0.10 (right).}
    \label{fig:random_walk_paths}
\end{figure}

\begin{figure}[t]
    \centering
    \includegraphics[width=0.5\linewidth]{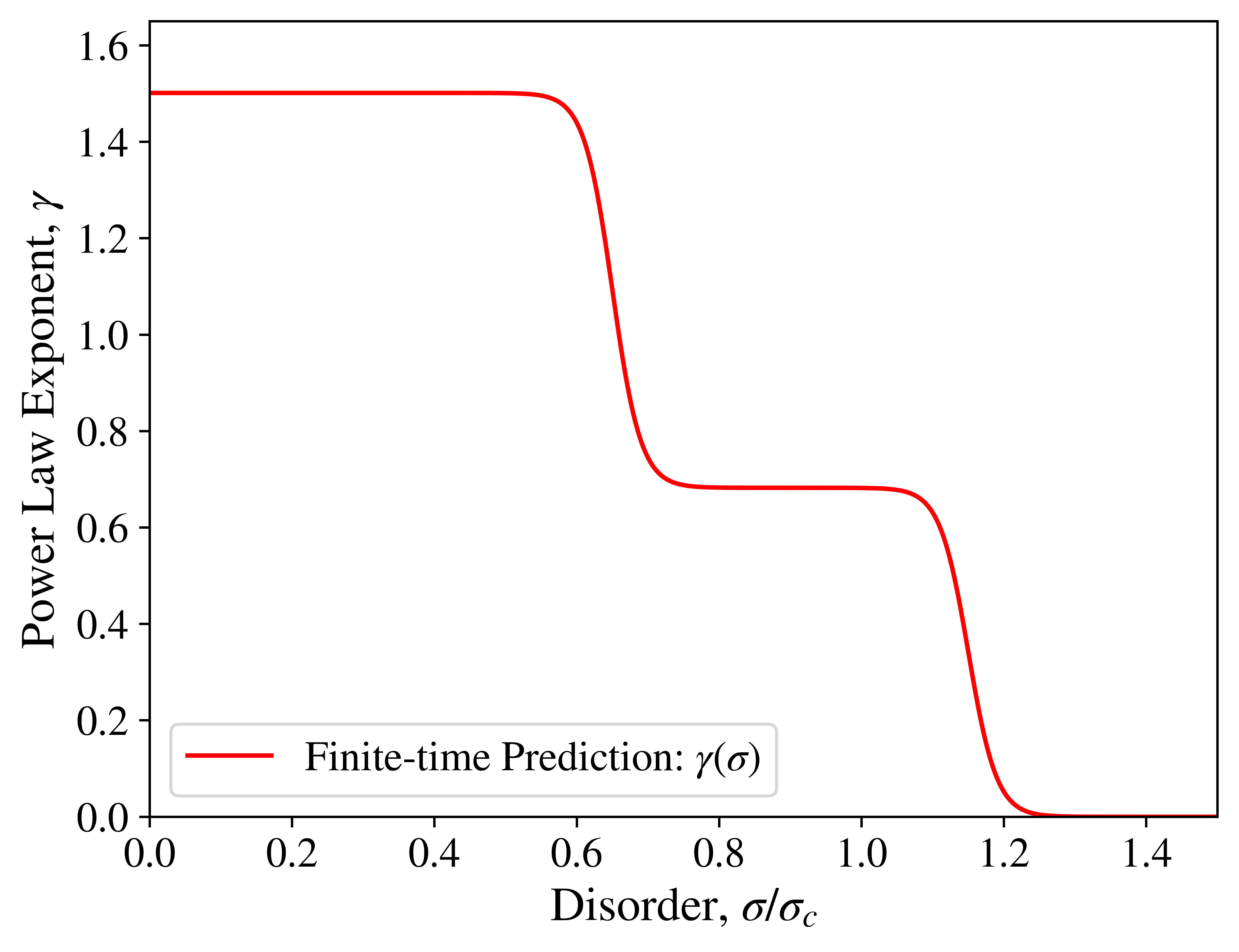}
    \caption{Finite time prediction of the rate of transport of magnetization on a disordered lattice, encompassing the diffusive, anomalous, and localized regimes predicted theoretically and verified with numerical simulations.}
    \label{fig:finite_t_prediction}
\end{figure}

Historically, anomalous behavior in transport properties of microscopic materials was first noticed in the electronic spin-relaxation of proteins \cite{Stapleton1980}.  Folded proteins can have a structure analogous to the path of a self-avoiding random walk, a fractal structure. These early ESR measurements could be explained by modifying the Raman relaxation mechanism with an anomalous vibrational density of states, inherited from the fractal dimension of the underlying protein structure. These early experimental results were quickly generalized by Alexander and Orbach to general fractals, introducing the notion of the \textit{fracton dimension}, which they postulated was equal to $4/3$ for percolation clusters of any dimension \cite{Alexander1982}. Anomalous diffusion is expected for random walkers on a fractal lattice \cite{Argyrakis1985, Gefen1982}, since the mean-squared displacement of a random walk inherits the fractal's geometric scaling, measured through $\bar{\delta}>0$:
\begin{equation}
    \avg{R^2(t)} \sim t^{\frac{2}{2+\bar{\delta}}}.
\end{equation}
In $d$ Euclidean dimensions, volume scales as $r^d$. For a fractal, the volume scaling is anomalous, with dimension $\bar{d}$. Generically, one expects that for aperiodic random walks, the survival probability scales inversely with the number of accessible sites, e.g. the accessible volume (at time $t$). The survival probability is thus expected to decay with rate depending on both $\bar{d}$ and $\bar{\delta}$, such that
\begin{equation}
    P_0(t) \sim 1/\avg{R^2(t)}^{\bar{d}/2} \sim t^{-\frac{\bar{d}}{2+\bar{\delta}}} = t^{-\bar{\bar{d}}/2}.
\end{equation}
The power law decay of the survival probability is then given by the \textit{fracton dimension} of the fractal,
\begin{equation}
    \bar{\bar{d}} = \frac{2\bar{d}}{2+\bar{\delta}},
\end{equation}
which is approximately $4/3$ for percolation clusters in all dimensions \cite{Alexander1982}. In panel (b) of Figure \ref{fig:msd_disorder}, the anomalous region of the mean-squared displacement curve closely agrees with the expectations of a random walker on a percolation fractal. In Figure \ref{fig:random_walk_paths}, we show a few random walk paths in different percolation regimes for the FCC system. In the rightmost panel, we plot three examples of random walks in a fully connected lattice with 1000 time steps taken in each trace. In the center panel, we show two traces of random walkers at/near the percolation phase transition. These traces show much more jagged dead-ends and a much smaller overall extent of motion, even though the simulation ran for 10,000 time steps. Finally, in the leftmost panel, we notice that below the percolation phase transition all subgraphs are finite; random walkers are unable to escape their local vicinity and will thermalize slowly to the stationary distribution on the locally finite graph, repeatedly tracing out its structure. Intuitively, since $D \propto \avg{R^2}/\avg{\tau}$, the system should only fail to be diffusive when either $\avg{R^2}\longrightarrow\infty$ (super-diffusion) or $\avg{\tau}\longrightarrow\infty$ (sub-diffusion). For percolation networks, subdiffusive behavior occurs since the fractal structure allows random walkers to take long detours that result in dead-ends -- these long dead-end paths work like deep trapping potentials on large scales and induce anomalous diffusion \cite{Bouchad1989}. Finally, in Figure \ref{fig:finite_t_prediction}, we include the finite time anomalous diffusion in our theoretical sketch for $\gamma(\sigma)$ -- Our theoretical measurements span a little more than a decade of time, over which time anomalous diffusion may persist for a wide range of disordered field strengths.  Classically, on a perfectly quenched percolation network, we would predict a power-law decay of $t^{-2/3}$, much slower than seen in the experimental data. As discussed in the main text, the fact that the observed power-law is faster than the Alexander and Orbach prediction is not surprising. Our data is still strong evidence that some fractal regime does emerge in the effective spin-lattice with a much faster emergent power law due to some combination of disorder correlations, long-range hopping, and bond regeneration.

\bibliography{SDPrefs}